\documentclass[%
 reprint,
 superscriptaddress,
 amsmath,amssymb,
 aps,
 prb,
]{revtex4-1}

\usepackage{graphicx}
\usepackage{dcolumn}
\usepackage{bm}

\begin{document}

\preprint{APS/123-QED}

\title{Frenkel pair formation energy for cubic Fe$_3$O$_4$ in DFT+U calculations}

\author{M.~I.~Shutikova}
    \email[]{shutikova.mi@phystech.edu}
    \affiliation{Joint Institute for High Temperatures of the Russian Academy of Sciences, Izhorskaya 13 Building 2, Moscow 125412, Russian Federation}
    \affiliation{Moscow Institute of Physics and Technologies (National Research University), Institutskij pereulok 9, Dolgoprudny Moscow region 141700, Russian Federation}
\author{V.~V.~Stegailov}
    \email[]{stegailov.vv@mipt.ru}
    \affiliation{Joint Institute for High Temperatures of the Russian Academy of Sciences, Izhorskaya 13 Building 2, Moscow 125412, Russian Federation}
    \affiliation{Moscow Institute of Physics and Technologies (National Research University), Institutskij pereulok 9, Dolgoprudny Moscow region 141700, Russian Federation}
    \affiliation{HSE University, Myasnitskaya ulitsa 20, Moscow 101000 Russian Federation}




\date{\today}

\begin{abstract}
Ab initio modelling of point defects in the cubic phase of magnetite faces two problems: the cubic structure becomes unstable below the Verwey temperature and there is no consensus on the electronic structure of the cubic phase (if there is a band gap, what type of symmetry of the wavefunction should be considered and how to describe strong electronic correlations). In this paper, we show that the comparison of the experimental data on the band gap and the Frenkel pair formation energy with the first-principles calculations allows to determine a consistent DFT+U model of cubic Fe$_3$O$_4$. 
\end{abstract}

\maketitle



	\textit{Introduction.} Being among the oldest materials known to the mankind, magnetite is still not well understood in the framework of the solid state theory. At the Verwey transition temperature $T_{\mathrm{V}} \sim 125$~K magnetite shows a sudden rise in conductivity and transforms from the low temperature monoclinic phase into the high temperature inverse spinel (cubic) phase~\cite{verwey1939electronic}. Magnetite exhibits ferrimagnetic ordering below 858~K and becomes paramegnetic at higher temperatures. 
	
	The puzzle of the Verwey transition is an important topic in the physics of strongly correlated systems and attracts a lot of attention~\cite{walz2002verwey,garcia2004verwey}. Following Vervey, this transition can be described within the order-disorder formalism. Charge and orbital order mechanisms~\cite{kugel1982jahn,streltsov2017orbital} are been considered~\cite{leonov2004charge,attfield2015orbital,gastaldo2021direct}. Recently, a new type of excitations called trimerons were identified in the low temperature phase~\cite{senn2012charge}. The experiments, however, provide controversial evidences on the possibility of trimeron correlations above the Verwey temperature~\cite{baldini2020discovery,elnaggar2020possible}. The recent core-level x-ray spectroscopy measurements give new information on the cation ordering at temperatures up to 1200~K~\cite{elnaggar2021temperature} including the formation of cation Frenkel pairs. This experiment sheds new light on the previous thorough studies of high temperature properties of magnetite by Dieckmann~et al.~\cite{dieckmann_defects_1977,dieckmann_defects_1977-1,dieckmann_defects_1978,dieckmann_defects_1982,dieckmann_defects_1983,dieckmann_defects_1986,dieckmann_point_1998} and emphasize the role of point defects in understanding the properties of magnetite at high temperatures.
	
	One of the results of the extensive experimental studies of Dieckmann and coauthors is the value of the cation Frenkel pair formation energy in magnetite ${E_{\mathrm{FP}}^\mathrm{f\ exp}=1.38}$~eV at about 900~K~\cite{dieckmann_defects_1986}. Although some ab initio data concerning magnetite point defects have been obtained previously~\cite{hendy_ab_2003,arras_electronic_2013,muhich_first-principles_2016,li_influence_2016,lininger_energetics_2018}, there is still not a single comparison between theoretical and experimental formation energies for defects in magnetite. In this work, we present the results of the density functional theory calculations with the Hubbard correction (DFT+U~\cite{anisimov1993density,anisimov_first-principles_1997}) for the formation energies of vacancies and interstitials and compare the results with $E_{\mathrm{FP}}^\mathrm{f\ exp}$. Following Liu and Di~Valentin~\cite{liu_band_2017} and our recent study~\cite{shutikova_vacancy_2021}, we use the static DFT+U model of cubic magnetite without symmetry constraints on electron density that predicts the existence of a small band gap above $T_{\mathrm{V}}$ that is in agreement with several experiments~\cite{park_single-particle_1997,gasparov_electronic_2007,banerjee2019track}.
	
\begin{figure*}
    \begin{minipage}[]{0.95\linewidth}
			\begin{center}
				\includegraphics[width=0.49\linewidth]{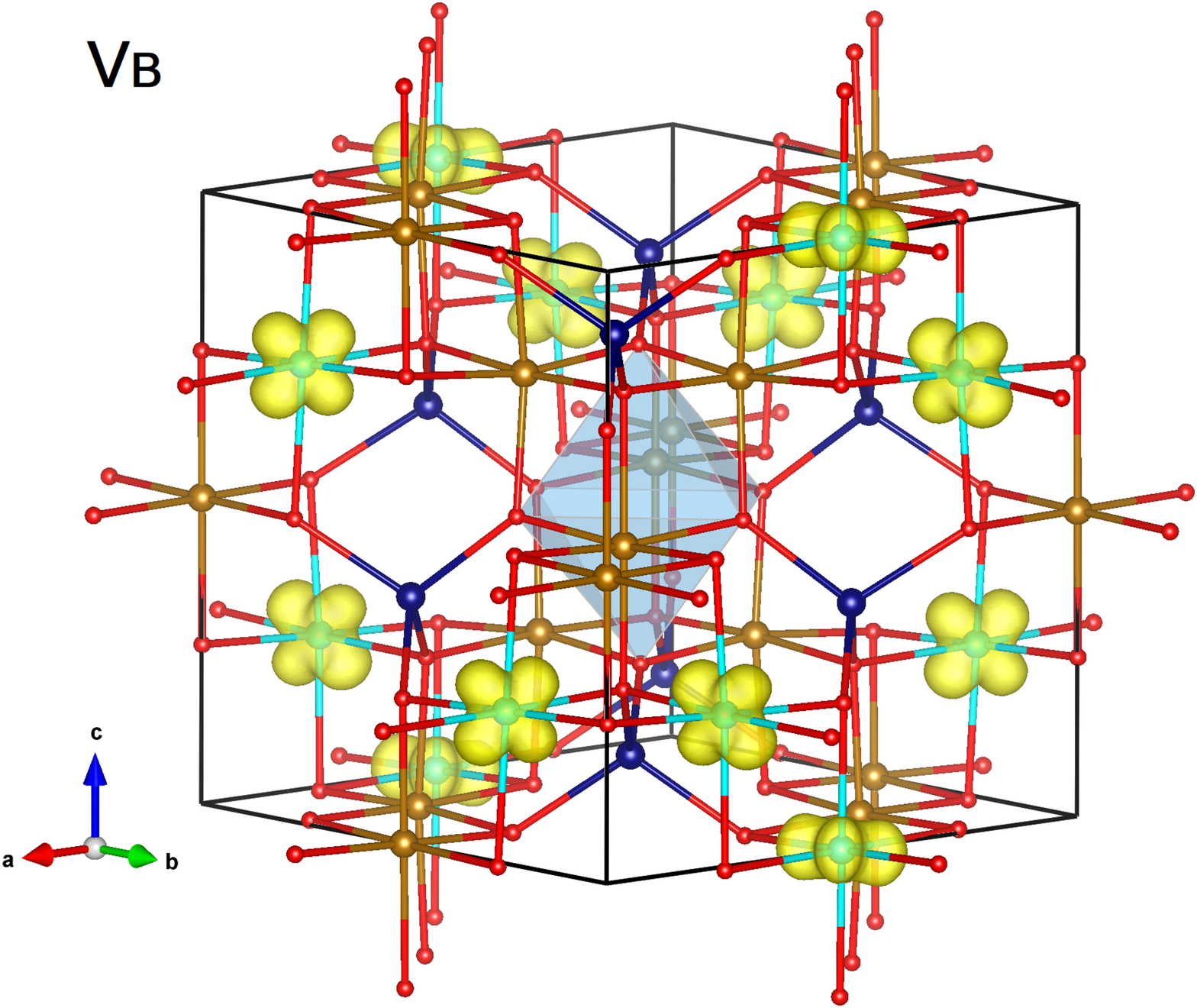}
				\includegraphics[width=0.49\linewidth]{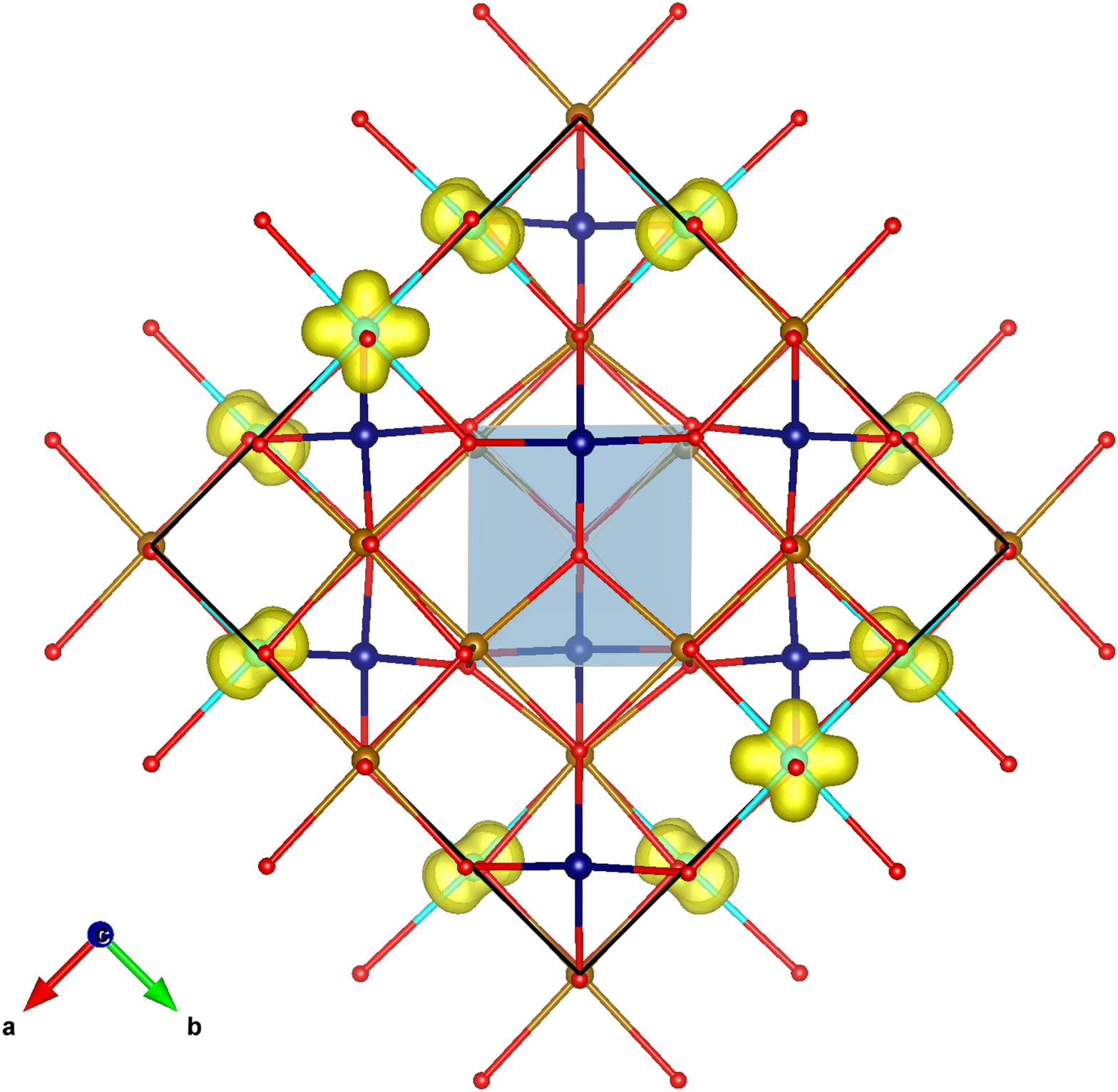}
			\end{center}
		\end{minipage}
		\begin{minipage}[]{0.95\linewidth}
			\begin{center}
				\includegraphics[width=0.49\linewidth]{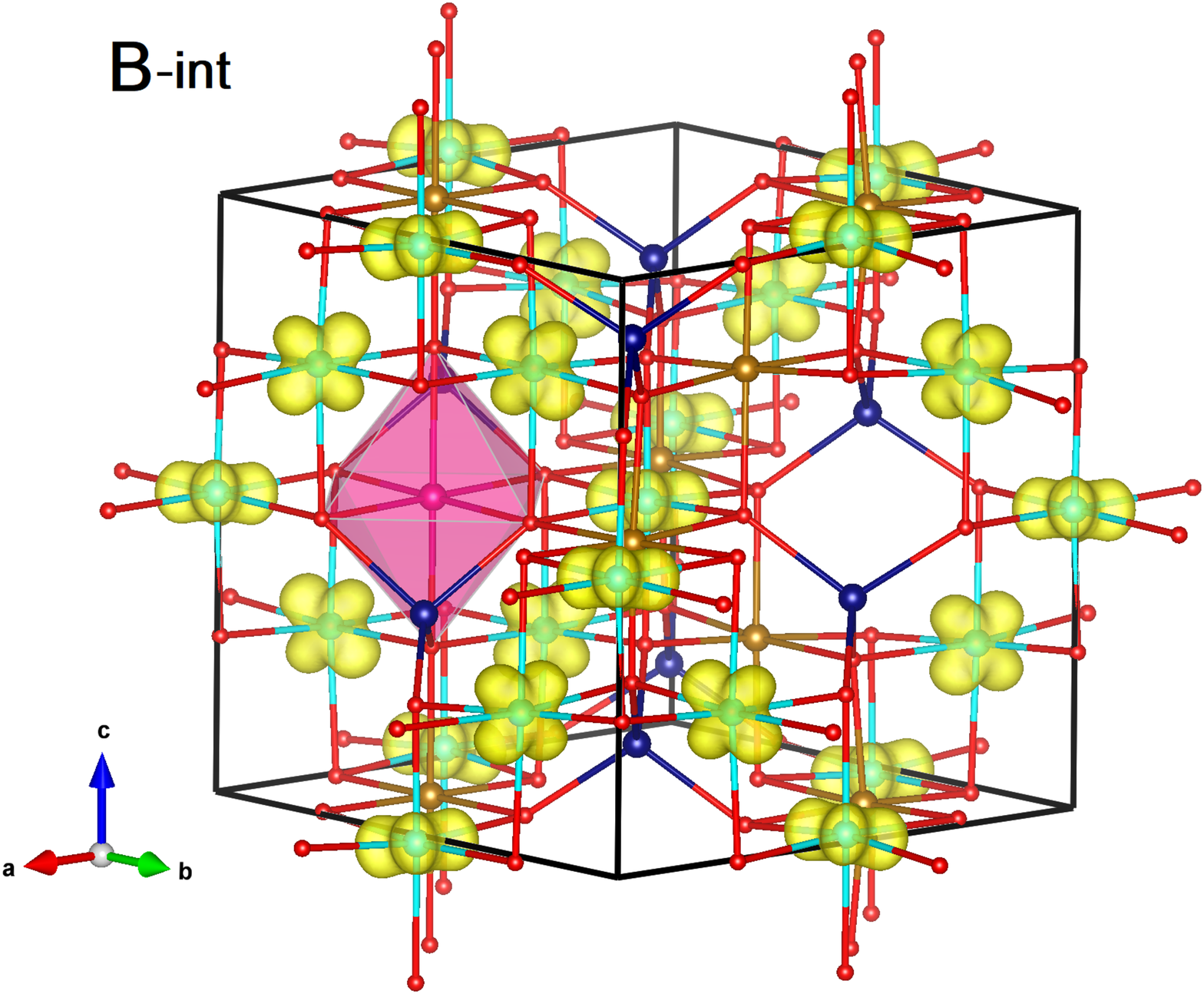}
				\includegraphics[width=0.49\linewidth]{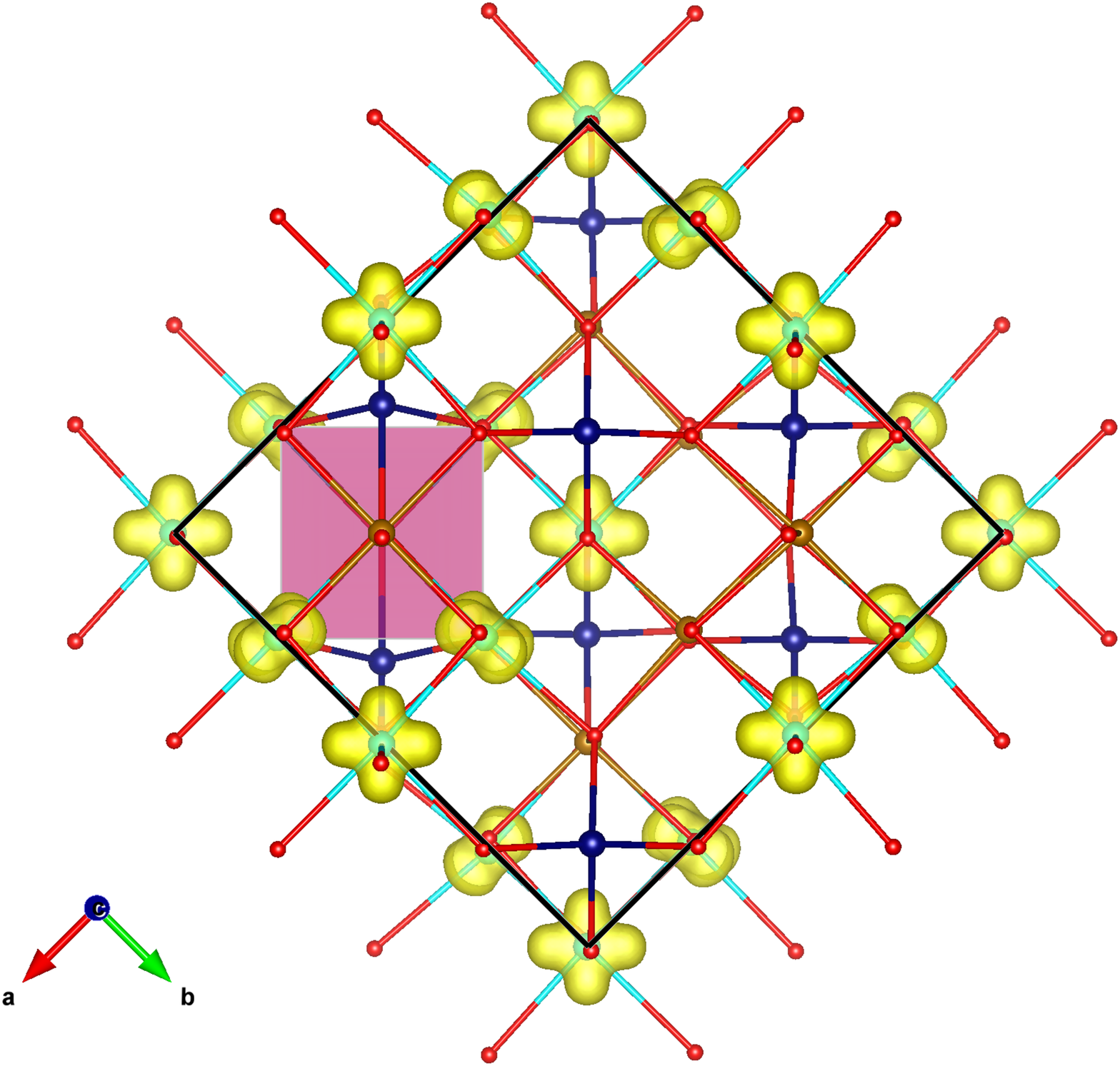}
			\end{center}
		\end{minipage}
    \caption{\label{fig:def}The optimized geometry with partial spin density for octahedral vacancy (V$_\mathrm{B}$) and octahedral interstitial atom (B-int) calculated with $U_{\mathrm{eff}}=3.65$~eV. VESTA program \cite{momma_vesta_2011} is used for visualization. Fe$_\mathrm{A}^{3+}$, Fe$_\mathrm{B}^{2+}$, Fe$_\mathrm{B}^{3+}$, and O$^{2-}$ are shown as dark-blue, ligth-blue, gold, and red spheres. The light-blue octahedron of the B-vacancy and pink polyhedron of the self-interstitial atom are shown.}
    \end{figure*}

	
	A deviation from ideal stoichiometry at high oxygen potentials may be achieved by increasing the number of vacancies in the octahedral sublattice (B-vacancies, $\mathrm{V}_\mathrm{B}$, Fig.~\ref{fig:def}) whereas the electroneutrality is maintained by increasing the number of trivalent ions. A deviation from stoichiometry at low oxygen potentials may be achieved by increasing the number of divalent cations in octahedral interstitals ($\mathrm{B}_{\mathrm{int}}$--positions, Fig.~\ref{fig:def}), which are free in the ideal cubic phase~\cite{sundman_assessment_1991,hallstrom_modeling_2011}. Our earlier results~\cite{shutikova_vacancy_2021} confirm that the formation energy of a vacancy in the B-sublattice is lower than in the A- (or tetrahedral) sublattice. Calculations of the formation energies of iron interstitial atoms in the DFT+U framework is one of the tasks of this study.

	The use of the symmetry constraint on the electron density and the wave function in the DFT+U model of defect-free magnetite cubic phase is a problem of current interest~\cite{liu_band_2017,shutikova_vacancy_2021}. It goes without saying that for defect magnetite such problem is not actual: a defect breaks lattice symmetry. However, the total energy of a defect-free configuration $E_0$ is needed to estimate the defect formation energy, and ${E_0^{\mathrm{sym}}\neq E_0^{\mathrm{asym}}}$ for models with and without the symmetry constraint respectively~\cite{liu_band_2017}. 
	
	It was shown that the asymmetric ground state of cubic Fe$_3$O$_4$ has a lower total energy than the symmetric ground state for DFT+U calculations even though the geometry has a cubic phase symmetry in both cases~\cite{liu_band_2017}. At a sufficiently high Hubbard parameter $U_{\mathrm{eff}}$ the asymmetric ground state gives the differences between di- and trivalent B-cations and a band gap that is confirmed in calculations with hybrid functionals~\cite{liu_band_2017}. At the same time, no differences between di- and trivalent B-cations and no band gap are found in other recent DFT+U studies of the magnetite cubic phase~\cite{li_influence_2016,piekarz_trimeron-phonon_2021}. This may indicate that the ground state obtained within the symmetric ansatz for electronic structure was used in these studies (the authors did not provide the details).
	
	There are three main goals of this study: 1) to find the iron interstitial atom configuration with the lowest formation energy, 2) to calculate Frenkel pair formation energies for different values of $U_{\mathrm{eff}}$, and 3) to refine the magnetite cubic phase model by comparing the Frenkel pair formation energy and the band gap calculated in the DFT+U framework with experimental data.
	

\textit{Calculation details.} All calculations in the study are carried out in the framework of either the spin polarized density functional theory (DFT) or DFT with the Hubbard correction term taking into account strong electronic correlations (the DFT+U method~\cite{anisimov1993density,anisimov_first-principles_1997}). The DFT calculations are performed in VASP~\cite{kresse_ab_1993,kresse_efficient_1996} with the PAW models for Fe and O and the PBE exchange-correlation functional. The Dudarev DFT+U scheme with ${U_{\mathrm{eff}}=U-J}$ is used \cite{dudarev_electron-energy-loss_1998}. The energy cutoff of 550~eV is used for the plane wave basis set. The $\Gamma$-centered k-point grids ${6\times6\times6}$, ${5\times5\times5}$, and ${3\times3\times3}$ are used for the supercells with 56, 112, and 448 atoms respectively. The total energy convergence threshold for self-consistent calculations and the forces-on-atoms convergence threshold for geometry optimizations are $10^{-6}$~eV and $10^{-2}$~eV/$\AA$ respectively. The geometry optimization of the supercells with a defect and the atomic relaxation of defect-free supercells are performed at a fixed equilibrium lattice constant~\cite{hendy_ab_2003, arras_electronic_2013, shutikova_vacancy_2021}. Equilibrium (zero pressure) lattice constants are obtained for the cubic phase of Fe$_3$O$_4$ with the oxygen parameter $x=0.2549$ (pressure has a strong influence on magnetite structure~\cite{rozenberg1996nature,ding2008novel}). A single pattern of charge-orbital ordering is considered, which has the lowest total energy (the case 'm2 -545' in~\cite{shutikova_vacancy_2021}).
	
	The self-interstitial and the Frenkel pair formation energies are calculated as:
	\begin{equation}
		E^\mathrm{f}_\mathrm{I}=E_\mathrm{I}-E_0-E_{\mathrm{at}},
	\end{equation}
	\begin{equation}
		E_{\mathrm{FP}}^{\mathrm{f}}=E_{\mathrm{I}}+E_{\mathrm{V}}-2E_0,
		\label{FP}
	\end{equation} 
	where $E_{\mathrm{I}}$, $E_{\mathrm{V}}$, and $E_0$ are the total energies of the self-interstitial atom, the vacancy and the defect-free supercells respectively; $E_{\mathrm{at}}$ is the iron chemical potential (the Frenkel pair formation energy does not depend on $E_{\mathrm{at}}$).


\textit{Results.} Table~\ref{tabl_def} summarizes the results of this work. The models of the defect-free cubic phase obtained in DFT and DFT+U using cubic supercells with 56 atoms have been reported earlier~\cite{shutikova_vacancy_2021}, they are presented in Table~\ref{tabl_def} for comparison. The supercell size effect on magnetite properties is given using supercells with 112 and 448 atoms in the DFT and DFT+U  ($U_{\mathrm{eff}}=3.5$~eV) frameworks.
	
	The calculations with and without the symmetry constraint on the electron density and on the wave function are carried out in DFT and DFT+U. In pure DFT without taking into account strong electronic correlations (${U_{\mathrm{eff}}=0}$), there are no differences between asymmetric and symmetric ground states. However in DFT+U the asymmetric ground state has lower total energy than the symmetric one (the case denoted as \textit{sym} in Table~\ref{tabl_def}) that is an agreement with the previous results~\cite{liu_band_2017,shutikova_vacancy_2021}.
	
	The symmetric ground state in DFT+U has some properties similar to those obtained in DFT without the Hubbard U: there are no differences between di- and trivalent B-cations and no band gap. Also, there are no significant structural changes after atomic relaxation at a fixed lattice constant. However, the cation magnetic moments in the DFT+U symmetric case are higher than in DFT, and they are similar to the data reported recently~\cite{li_influence_2016,piekarz_trimeron-phonon_2021}.
	
\begin{figure}
		\center \includegraphics[width=0.9\linewidth]{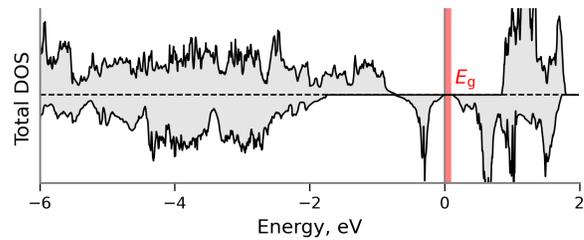}
		\caption{{The electron density-of-states (DOS) of both spin components for the cubic phase at $U_{\mathrm{eff}}=3.65$~eV. A band gap in the minor spin DOS is marked.}}
		\label{fig_Eg}
	\end{figure}

	The DFT+U asymmetric ground state of magnetite cubic phase shows the differences between di- and trivalent B-cations and a non-zero band gap~\cite{shutikova_vacancy_2021} (Fig.~\ref{fig_Eg}). The band gap width, the equilibrium lattice constant and the cation magnetic moments all depend of $U_{\mathrm{eff}}$. The plots of these dependencies are given in the our previous work~\cite{shutikova_vacancy_2021} and in the Supplementary Materials (SM). 

	\begin{table*}
    \caption{\label{tabl_def} The properties of the ideal magnetite defect-free cubic phase supercell: the number of atoms in the defect-free supercell (($N_{\mathrm{at}}$), the optimized lattice constant ($a_0$, $\AA$), the total energy per formula unit ($E_0$, eV/f.u.), the magnetic moments ($\mu_{{\mathrm{Fe}}_{\mathrm{A}}}$, $\mu_{{\mathrm{Fe}}_{\mathrm{B}}^{3+}}$, $\mu_{{\mathrm{Fe}}_{\mathrm{B}}^{2+}}$, $\mu_{\mathrm{B}}$), the band gap ($E_{\mathrm{g}}$, meV), and the total external pressure ($p_0$, kbar). The properties after atomic relaxation of defect-free supercell at a fixed lattice constant $a_0$: the total energy difference ($\Delta E^{\mathrm{rel}}=E^{\mathrm{rel}}-E_0$, eV/f.u.), the band gap ($E_{\mathrm{g}}^{\mathrm{rel}}$, meV), and the total external pressure ($p$, kbar). The properties of defect supercell with a vacancy or an interstitial after atomic relaxation at a fixed lattice constant $a_0$: the total energy of defect supercell ($E_{\mathrm{def}}$, eV/f.u.), the total energy difference ($\Delta E^{\mathrm{def}}=E^{\mathrm{def}}-E_0$, eV), the chemical potential of iron ($E_{\mathrm{at}}$, eV), the defect formation energies ($E_{\mathrm{def}}^{\mathrm{f}}$, $E_{\mathrm{FP}}^\mathrm{f}$, eV), the total magnetic moment change ($\Delta \mu=\mu_{\mathrm{tot}}^{\mathrm{def}}-\mu_{\mathrm{tot}}^0$, $\mu_{\mathrm{B}}$), the magnetic moment of the self-interstitial iron ($\mu_{{\mathrm{Fe}}_{\mathrm{int}}}$,  $\mu_{\mathrm{B}}$).}
    \begin{ruledtabular}
        \begin{tabular}{c|cc|c|c|ccc|ccc}
					model              & \multicolumn{2}{c|}{DFT} & \multicolumn{7}{c}{DFT+U} \\
					$U_{\mathrm{eff}}$          & \multicolumn{2}{c|}{0} & $ 3.40$   & \multicolumn{4}{c|}{$3.50$} & $3.60$ & $3.65$ & $3.84$   \\
					$N_{\mathrm{at}}$           & $56$ & $448$ & $56$   & $56\ \mathrm{sym}$         & $56$      & $112$     & $448$     & $56$   & $56$  & $56$       \\
					\hline
					\multicolumn{11}{c}{defect-free bulk} \\
					\hline
					$a_0$ 		       & $8.392$ & $8.392$ & $8.478$ & $8.456$ & \multicolumn{3}{c|}{$8.479$}     & $8.4800$       & $8.4805$  & $8.483$      \\
					$E_0$        & $-53.449$ & $-53.450$ & $-48.711$   & $-48.319$          & $-48.605$      & $-48.606$     & $-48.607$     & $-48.502$      & $-48.454$     & $-48.264$   \\
					$\mu_{{\mathrm{Fe}}_{\mathrm{A}}}$       & $-3.48$ & $-3.48$ & $-4.02$   & $-4.03$          & $-4.02$      & $-4.02$     & $-4.02$     & $-4.03$      & $-4.04$   & $-4.06$    \\
					$\mu_{{\mathrm{Fe}}_{\mathrm{B}}^{2+}}$  & $3.56$ & $3.56$ & $3.68$   & $3.93$          & $3.68$      & $3.68$     & $3.68$     & $3.68$      & $3.68$  & $3.68$       \\
					$\mu_{{\mathrm{Fe}}_{\mathrm{B}}^{3+}}$  & $3.56$ & $3.56$ & $4.10$   & $3.93$          & $4.11$      & $4.11$     & $4.11$     & $4.12$      & $4.13$  & $4.15$      \\
					$E_{\mathrm{g}}$              & \multicolumn{2}{c|}{no gap} & $39$   & no gap          & $61$      & $12$     & $62$     & $86$      & $105$ & $155$ \\
					$p_0$                & $0.7$ & $0.1$ & $-0.3$   & $0.0$          & $0.0$      & $0.8$     & $0.3$     & $0.1$      & $0.1$ & $-0.1$ \\			
					\hline
					\multicolumn{11}{c}{after atomic relaxation} \\
					\hline
					$\Delta E^{\mathrm{rel}}$ & $0.000$ & $-0.001$ & $-0.104$   & $-0.001$          & $-0.106$      & $-0.082$     & $-0.105$     & $-0.107$      & $-0.108$ & $-0.110$       \\
					$E_{\mathrm{g}}^{\mathrm{rel}}$        & \multicolumn{2}{c|}{no gap} & $572$   & no gap      & $588$      & $360$     & $583$     & $636$      & $660$  & $860$      \\
					$p$         & $0.0$ & $-0.7$ & $1.9$   & $0.0$          & $2.3$      & $5.1$     & $1.9$     & $2.7$      & $2.8$ & $3.0$ \\	
					\hline
					\multicolumn{11}{c}{B-vacancy} \\
					\hline
					$E_{\mathrm{def}}$          & $-52.291$ & $-53.307$ & $-47.860$   & $-47.748$          & $-47.759$      & $-48.210$     & $-48.587$     & $-47.659$      & $-47.614$   & $-47.431$     \\
					$\Delta E^{\mathrm{def}}$      & $9.260$ & $9.206$ & $6.813$   & $4.566$          & $6.773$      & $6.342$     & $1.297$     & $6.739$      & $6.723$    & $6.666$    \\
					$E_{\mathrm{at}}$           & \multicolumn{2}{c|}{$-8.238$} & $-5.639$   & \multicolumn{4}{c|}{$-5.571$}     & $-5.504$      & $-5.473$     &  $-5.348$ \\
					$E_{\mathrm{def}}^{\mathrm{f}}$        & $1.02$ & $0.97$ & $1.17$   & $-1.01$          & $1.20$      & $0.77$     & $-4.27$     & $1.24$      & $1.25$    &  $1.32$  \\
					$\Delta \mu$       & $-2$ & $-2$ & $-2$   & $-2$          & $-2$      & $-2$     & $-2$     & $-2$      & $-2$    & $-2$    \\
					$p$         & $-5.9$ & $-1.5$ & $-10.5$   & $2.5$          & $-10.6$      & $-5.5$     & $0.0$     & $-10.6$      & $-10.6$ & $-11.0$\\		
					\hline
					\multicolumn{11}{c}{B-interstitial} \\
					\hline
					$E_{\mathrm{def}}$          & $-54.086$ & $-$ & $-49.364$   & $-49.221$          & $-49.260$      & $-48.940$     & $-48.782$     & $-49.157$      & $-49.111$  & $-48.923$      \\
					$\Delta E^{\mathrm{def}}$      & $-5.098$ & $-$ & $-5.221$   & $-7.220$          & $-5.235$      & $-5.351$     & $-11.164$     & $-5.245$      & $-5.250$   & $-5.269$     \\
					$E_{\mathrm{def}}^{\mathrm{f}}$        & $3.14$ & $-$ & $0.42$   & $-1.65$          & $0.34$      & $0.22$     & $-5.59$     & $0.26$      & $0.22$  &  $0.08$    \\
					$\Delta \mu$       & $-6$ & $-$ & $-6$   & $-6$          & $-6$      & $2$     & $-2$     & $-6$      & $-6$  & $-6$      \\
					$\mu_{{\mathrm{Fe}}_{\mathrm{int}}}$        & $-3.41$ & $-$ & $-3.70$   & $-3.72$          & $-3.70$      & $3.59$     & $-3.67$     & $-3.71$      & $-3.71$     & $-3.72$   \\
					$p$         	   & $33.4$ & $-$ & $38.7$   & $55.7$          & $38.8$      & $21.7$     & $7.1$     & $38.9$      & $39.0$ & $38.8$ \\		
					\hline
					$E_{\mathrm{FP}}^\mathrm{f}$          & \textbf{4.16} & $-$ & \textbf{1.59}  & \textbf{--2.65}          & \textbf{1.54}      & \textbf{0.99}     & \textbf{--9.87}     & \textbf{1.50}      & \textbf{1.47}     & \textbf{1.40}    \\
					\hline
					$E_{\mathrm{FP}}^\mathrm{f\ rel}$          & $4.16$ & $-$ & $3.26$   & $-2.64$          & $3.23$      & $3.61$     & $3.56$     & $3.20$      & $3.19$  &   $3.16$    \\
				\end{tabular}
        \end{ruledtabular}
    \end{table*}
	
	After atomic relaxation of a defect-free supercell at a fixed lattice constant the lattice symmetry is distorted. The band gap after the relaxation increases by an order of magnitude due to decrease in the degree of overlap of atomic orbitals that rotate during relaxation. The total energy of atomic structure after the relaxation is lower than the total energy for ideal cubic symmetry. Interestingly, the change in the total energy per formula unit is approximately the same in supercells with different sizes (Table~\ref{tabl_def}).
	
	A set of energy local minima can be obtained for the interstital configuration in DFT~\cite{hendy_ab_2003} and DFT+U~\cite{arras_electronic_2013} in calculations with different initial approximations to wavefunctions and spin density. In this work, we have found a lower B-vacancy formation energy (1.20~eV in Table~\ref{tabl_def}) than that was obtained in our previous study (1.26~eV in ~\cite{shutikova_vacancy_2021}).
	
	The local minima obtained for three initial geometries of iron interstitials in calculations with different initial values of the interstitial iron magnetic moment have been found (and discussed in detail in SM). The data collected in this work allow to conclude that the B-interstitial has the lowest formation energy among the variants of an isolated iron interstitial position in the cubic phase of magnetite. The solutions with the deepest minima for B-vac and B-int are presented in Table~\ref{tabl_def}. The optimized geometry and the partial spin density for B-interstitial with the lowest formation energy is shown in Fig.~\ref{fig:def} (see more details in SM).
	
	In DFT without the Hubbard correction the formation energy of the Frenkel pair is $4.16$~eV, which is in a fairly good agreement with the previous result $4.09$~eV~\cite{hendy_ab_2003}. However, this value is three times higher than the experimental value of the Frenkel pair formation energy ${E_{\mathrm{FP}}^\mathrm{f\ exp}=1.38}$~eV~\cite{dieckmann_defects_1986}.
	
	As a defect distorts the lattice symmetry, the calculations with initial symmetry constraint in defect supercells converge to results, which are similar to those without initial symmetry constraint (see $E_{\mathrm{def}}$ in Table \ref{tabl_def} for these cases). However, since ${E_0^{\mathrm{sym}}>E_0^{\mathrm{asym}}}$ in DFT+U, the formation energies of isolated defects and the Frenkel pair formation energy are negative for the case when the ground state of defect-free supercell is symmetric. This fact is an important argument against the applicability of the symmetric ansatz for the electronic structure of defect-free magnetite.
	
\begin{figure}
		\begin{minipage}[]{0.99\linewidth}
			\begin{center}
				\includegraphics[width=0.99\linewidth]{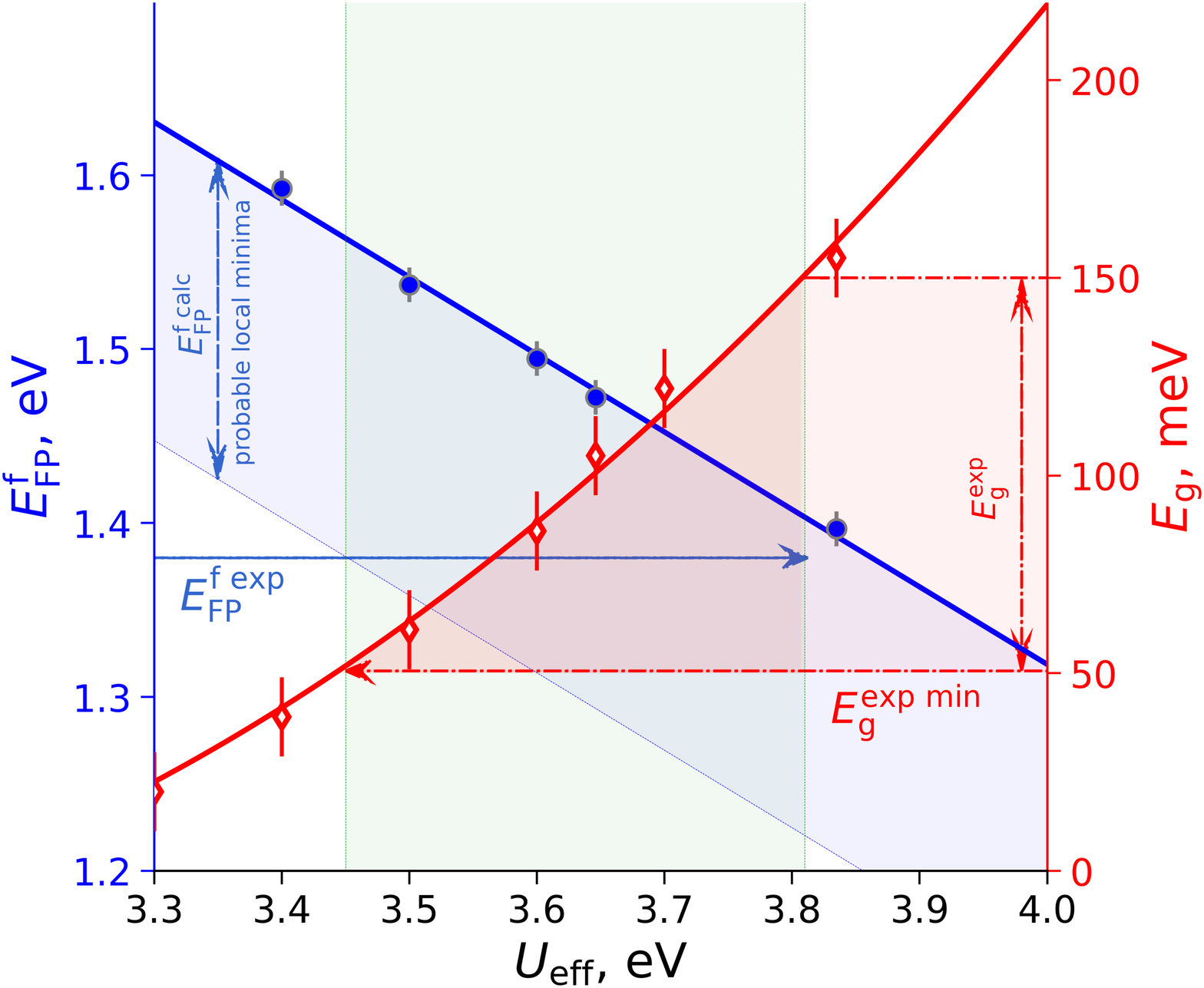}
			\end{center}
		\end{minipage}
		\caption{The effect of $U_{\mathrm{eff}}$ on the Frenkel pair formation energy and the band gap in magnetite cubic phase. The red fill illustrates the spread of the experimental results on the band gap $E_{\mathrm{g}}=100\pm50$~meV~\cite{park_single-particle_1997,gasparov_electronic_2007,banerjee2019track}, while the blue fill illustrates the hypothetical $E_{\mathrm{FP}}^\mathrm{f}$ values that could correspond to lower local energy minima or obtained (hypothetically) via QMD. The experimental value of $E_{\mathrm{FP}}^\mathrm{f}=1.38$~eV~\cite{dieckmann_defects_1986} is shown by the blue arrow. The $U_{\mathrm{eff}}$ optimum range is shown in green.}
		\label{fig_fp_Ueff}
	\end{figure}

	The influence of $U_{\mathrm{eff}}$ on the defect formation energies is shown in Fig.~\ref{fig_fp_Ueff}: with increasing $U_{\mathrm{eff}}$ the Frenkel pair formation energy decreases. The formation energy of B-vacancy (B-interstitial) increases (decreases) with increasing $U_{\mathrm{eff}}$ (see SM). The discussion is given below.
	
	The Frenkel pair formation energy calculated using cubic supercells containing 448 atoms is negative (Table~\ref{tabl_def}). This inadequate result should be attributed to more pronounced atomic relaxation in the large supercell. The total energy of a defect-free supercell with the atomic relaxation $E_0^{\mathrm{rel}}$ taken instead of $E_0$ in calculating the Frenkel pair formation energy (2) gives $E_{\mathrm{FP}}^\mathrm{f\ rel}$ (Table \ref{tabl_def}). These values are close to those obtained after the full geometry optimization in the earlier DFT+U study $E_{\mathrm{FP}}^\mathrm{f}=3.11$~eV~\cite{li_influence_2016}, so they could be interpreted as the defect formation energies in the monoclinic magnetite low temperature phase~\cite{pinto2006mechanism}.

    \textit{Discussion.} The equilibrium lattice constant of the cubic phase for different $U_{\mathrm{eff}}$ vary in the range ${8.478\div8.483\,\AA}$ that is 1\% larger the experimental value of ${8.385\,\AA}$~\cite{okudera1996temperature,levy2004effect}. The accuracy is rather high and is typical for DFT+U (e.g.~\cite{olsson2020stability}). Only the hybrid HSE06 functional was shown to give a better accuracy~\cite{liu_band_2017}.
	
	It is instructive to focus our attention on two other parameters that are more sensitive to $U_{\mathrm{eff}}$: $E_{\mathrm{g}}$ and $E_{\mathrm{FP}}^\mathrm{f}$.
	
	The experimental results~\cite{park_single-particle_1997,gasparov_electronic_2007,banerjee2019track} give the band gap width above $T_V$ in the range $50\div150$~meV (the most recent study gives $\sim60$~meV~\cite{banerjee2019track}). There are experimental evidences based on conductivity measurements that the band gap of magnetite cubic phase at elevated temperatures remains about 100~meV~\cite{nell1991high}. Therefore, the calibration by $E_{\mathrm{g}}$ gives the optimum range $U_{\mathrm{eff}} = {3.45\div3.81}$~eV. 
	
	The results on the Frenkel pair formation energy dependence on $U_{\mathrm{eff}}$ allow us to refine the model further. First of all, we see that zero band gap models of magnetite cubic phase above $T_V$ can not give the $E_{\mathrm{FP}}^\mathrm{f}$ values in a reasonable agreement with the experimental value ${E_{\mathrm{FP}}^\mathrm{f\ exp}=1.38}$~eV~\cite{dieckmann_defects_1986}. The best agreement of ${E_{\mathrm{FP}}^\mathrm{f\ exp}=1.38}$~eV with our DFT+U data on $E_{\mathrm{FP}}^\mathrm{f}$ gives $U_{\mathrm{eff}}$~=~3.86~eV that corresponds to $E_{\mathrm{g}}$ slightly above the experimental range. Here, we should note that the $E_{\mathrm{FP}}^\mathrm{f}$ values shown as blue points in Fig.~\ref{fig_fp_Ueff} correspond to the similar ionic structures and the same orbital/charge ordering patterns for varying $U_{\mathrm{eff}}$. There is a possibility that some deeper energy minima for a vacancy and/or for an interstitial could be found. Moreover, at finite $T > T_V$ defects exsist as dynamics structures and their energies of formation, strictly speaking, should take into account finite-T effects (e.g.~\cite{smirnov2019formation}). The corresponding quantum molecular dynamics (QMD) calculations are too computationally demanding and will not be able resolve this issue in the near future.
	
	We see that the use of a small supercell is a crucial condition to obtain an adequate agreement between the experimental and numerical data on the defect formation energies, because an accurate QMD modeling of cubic phase temperature stabilization in large supercells is a challenge at preset. In larger supercells, the band gap increase, the symmetry distortion and energy decrease are observed after the atomic relaxation. This relaxation reflects some features of the transition from the cubic phase to the monoclinic low-temperature phase in magnetite~\cite{pinto2006mechanism}.
	
	
	\textit{Conclusions.} Using the system size of 56 atoms, we compared the DFT+U solutions with the symmetrical ansatz for the wavefunction and the solutions without symmetry. In the former case, $E_{\mathrm{FP}}^\mathrm{f}$ is negative that disqualifies the symmetrical ansatz. In the latter case, $E_{\mathrm{FP}}^\mathrm{f}$ is positive that supports the asymmetrical wavefunction model.
	
	After careful selection of different interstitial configurations (and using the more refined results for vacancies than in~\cite{shutikova_vacancy_2021}), we have shown that DFT+U model of cubic Fe$_3$O$_4$ with $U_{\mathrm{eff}} = {3.45\div3.81}$~eV provides simultaneously the values of the band gap and values of the Frenkel pair formation energy in a reasonable agreement with experimental data. The cubic Fe$_3$O$_4$ models with lower $U_{\mathrm{eff}}$ (predicting a zero band gap) can not give an adequate values of $E_{\mathrm{FP}}^\mathrm{f}$.
	
	We have considered larger systems up to 448 atoms and showed that the static calculations for these systems including relaxation for the defect structure is unable to describe the experimental values of $E_{\mathrm{FP}}^\mathrm{f}$ for the cubic phase. The reason is the temperature stabilization of the cubic phase. Static relaxation of larger cells gives the defect structures that are closer to the low-temperature phase of magnetite below the Verwey temperature. The system of 56 atoms combines the relaxation of a point defect structure and the preservation of the cubic nature of the crystal matrix that effectively gives the best description for the point defects energies in the cubic phase within the static DFT+U calculations.
	
	
	\textit{Acknowledgements.} The study is supported by the Russian Foundation for Basic Research grants No.~20-21-00159 and No.~20-32-90081. The authors acknowledge the Supercomputer Centre of JIHT RAS and the Supercomputer Centre of MIPT. This research was supported in part through computational resources of HPC facilities at NRU HSE. This work was prepared in part within the framework of the HSE University Basic Research Program.

%

\end{document}


\maketitle
	
	\textbf{Defect-free cubic phase model}
	
\begin{figure}[h]
		\begin{minipage}[]{0.99\linewidth}
			\begin{center}
				\includegraphics[width=0.6\linewidth]{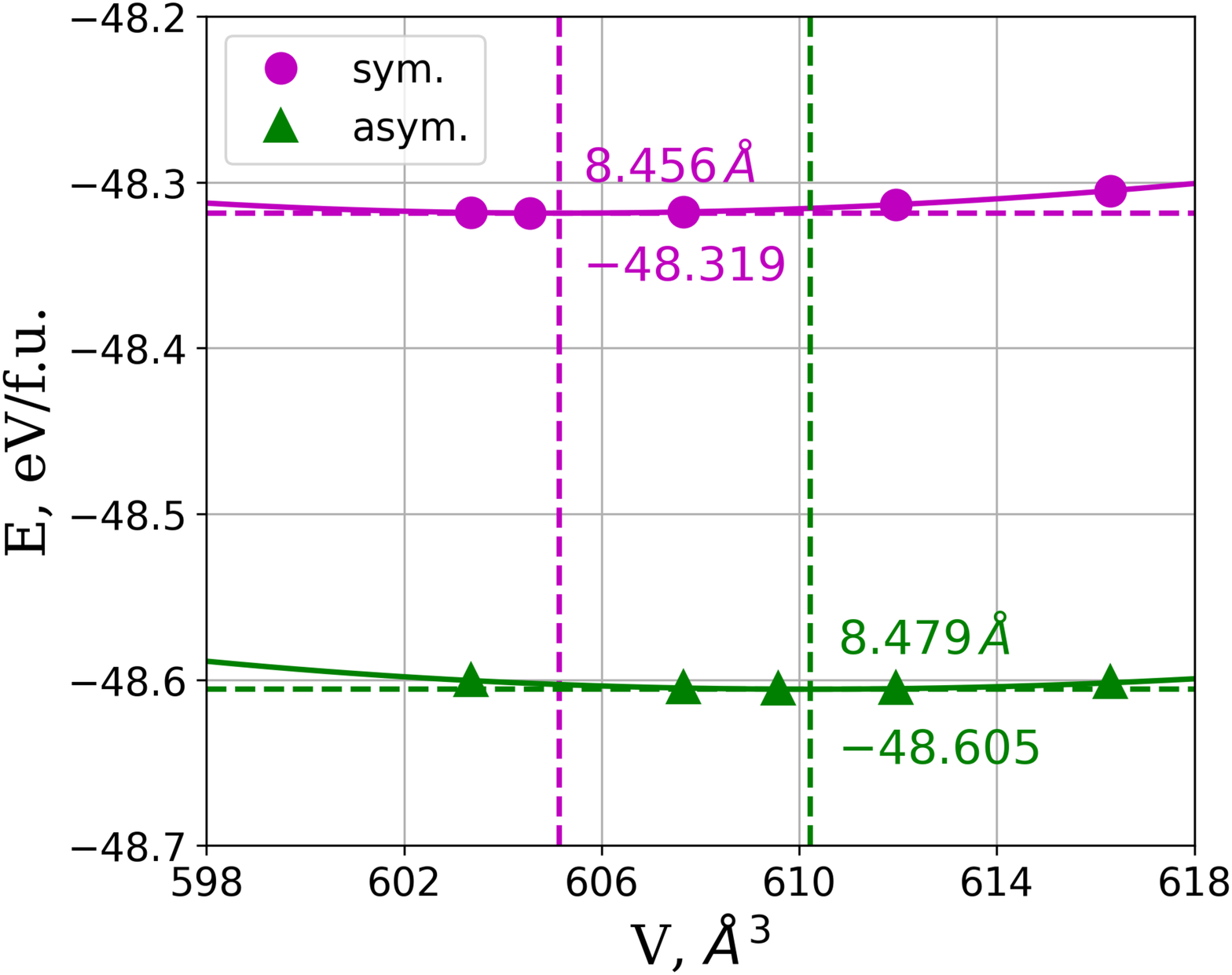}
			\end{center}
		\end{minipage}
		\caption{The total energies calculated at different lattice constants in cubic supercells with 56 atoms with and without the symmetry constraint on the electron density and on the wave function with ${U_{\mathrm{eff}}=3.5}$~eV. ${E_0^{\mathrm{sym}}=E_0^{\mathrm{asym}}+286}$~meV/f.u.}
		\label{fig_ae}
	\end{figure}

\begin{figure}[]
		\begin{minipage}[]{0.99\linewidth}
			\begin{center}
				\includegraphics[width=0.49\linewidth]{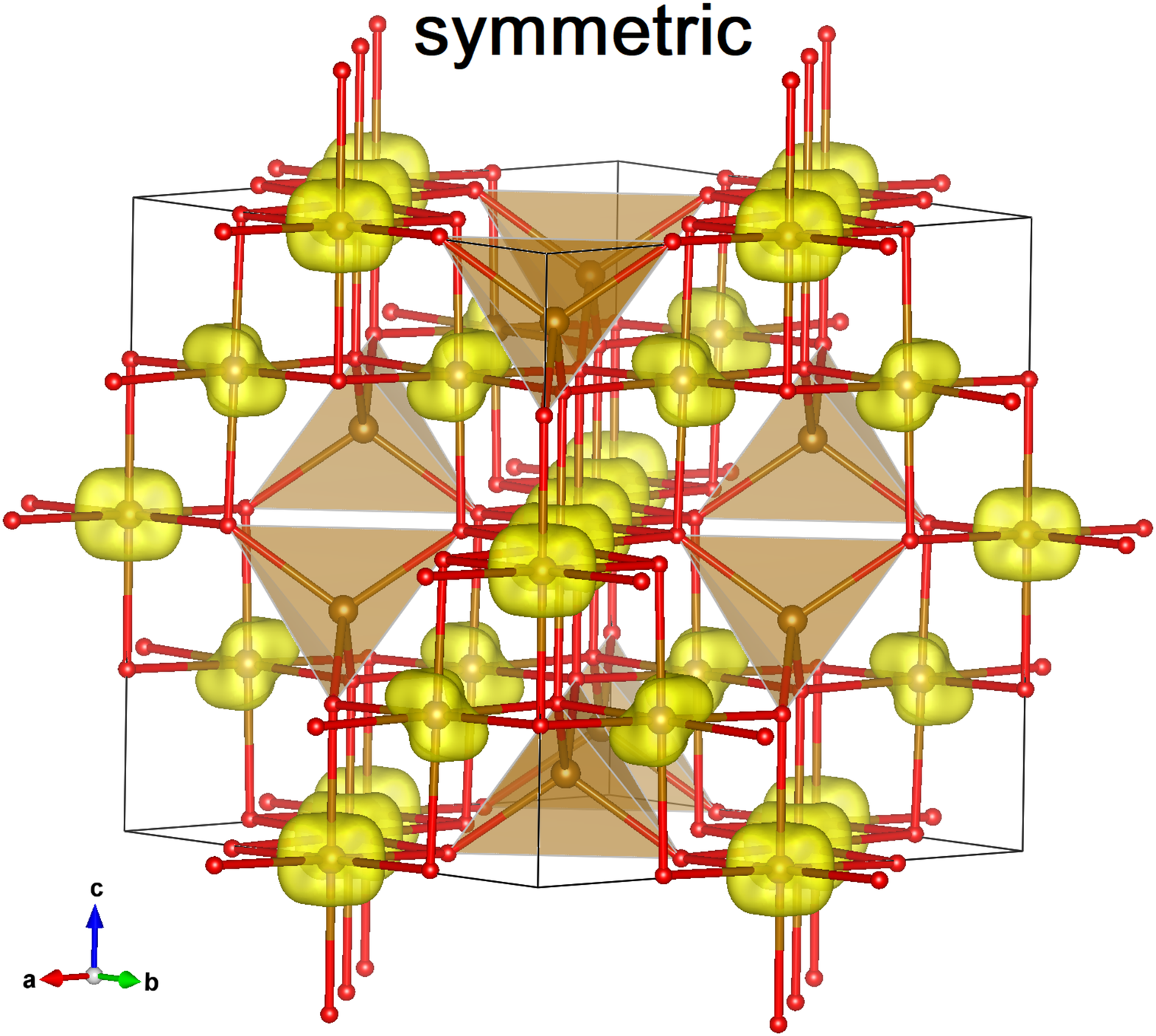}
				\includegraphics[width=0.45\linewidth]{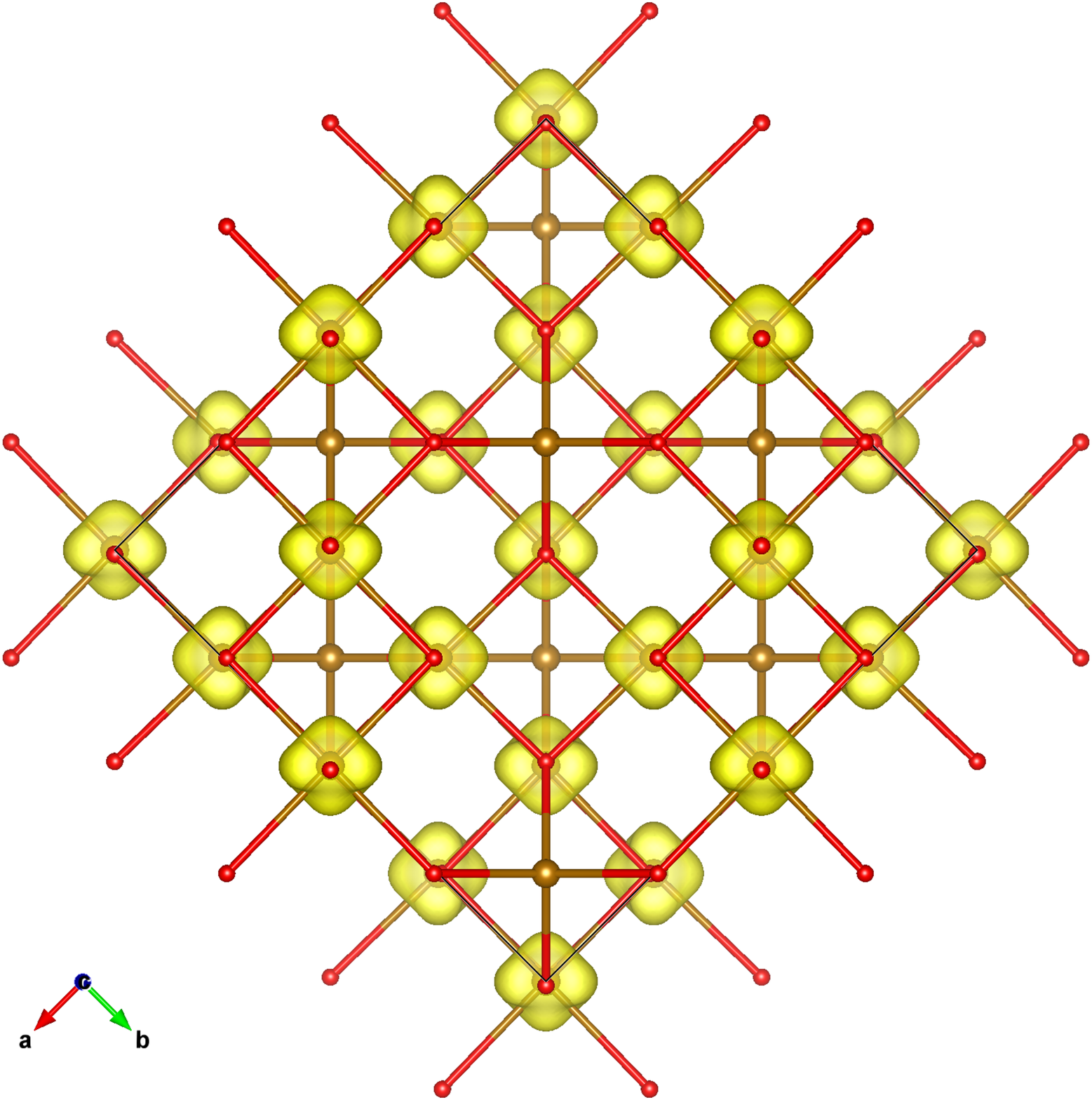}
			\end{center}
		\end{minipage}
		\caption{The partial spin density for bands in the range from $E_{\mathrm{F}}-0.75\,\mathrm{eV}$ to $E_{\mathrm{F}}$ for the symmetric ground state. VESTA program \cite{momma_vesta_2011} is used for visualization.}
		\label{fig_pcd_56}
	\end{figure}

	The total energies calculated at different lattice constants with and without the symmetry constraint on the electron density and on the wave function in DFT+U with ${U_{\mathrm{eff}}=3.5}$~eV are shown in Figure \ref{fig_ae}. The asymmetric ground state has indeed lower total energy than the symmetric one. The asymmetric ground state can be calculeted also with smaller ${U_{\mathrm{eff}}=3.2}$~eV\footnote{The calculation with ${U_{\mathrm{eff}}=3.2}$~eV gives $a_0=8.474\,\angstrom$, $\mu_{\mathrm{tot}}=4.0\,\mu_{\mathrm{B}}$, $\mu_{{\mathrm{Fe}}_{\mathrm{A}}}=-4.00\,\mu_{\mathrm{B}}$, $\mu_{{\mathrm{Fe}}_{\mathrm{B}}^{2+}}=+3.69\,\mu_{\mathrm{B}}$, $\mu_{{\mathrm{Fe}}_{\mathrm{B}}^{3+}}=+4.07\,\mu_{\mathrm{B}}$, $E_{\mathrm{g}}=10$~meV.}. The partial spin density for the symmetric ground state is shown in Figure \ref{fig_pcd_56} (see also the asymmetric case in Figure~11 of the paper~\cite{shutikova_vacancy_2021}).  
	
\begin{figure}[]
		\begin{minipage}[]{0.99\linewidth}
			\begin{center}
				\includegraphics[width=0.60\linewidth]{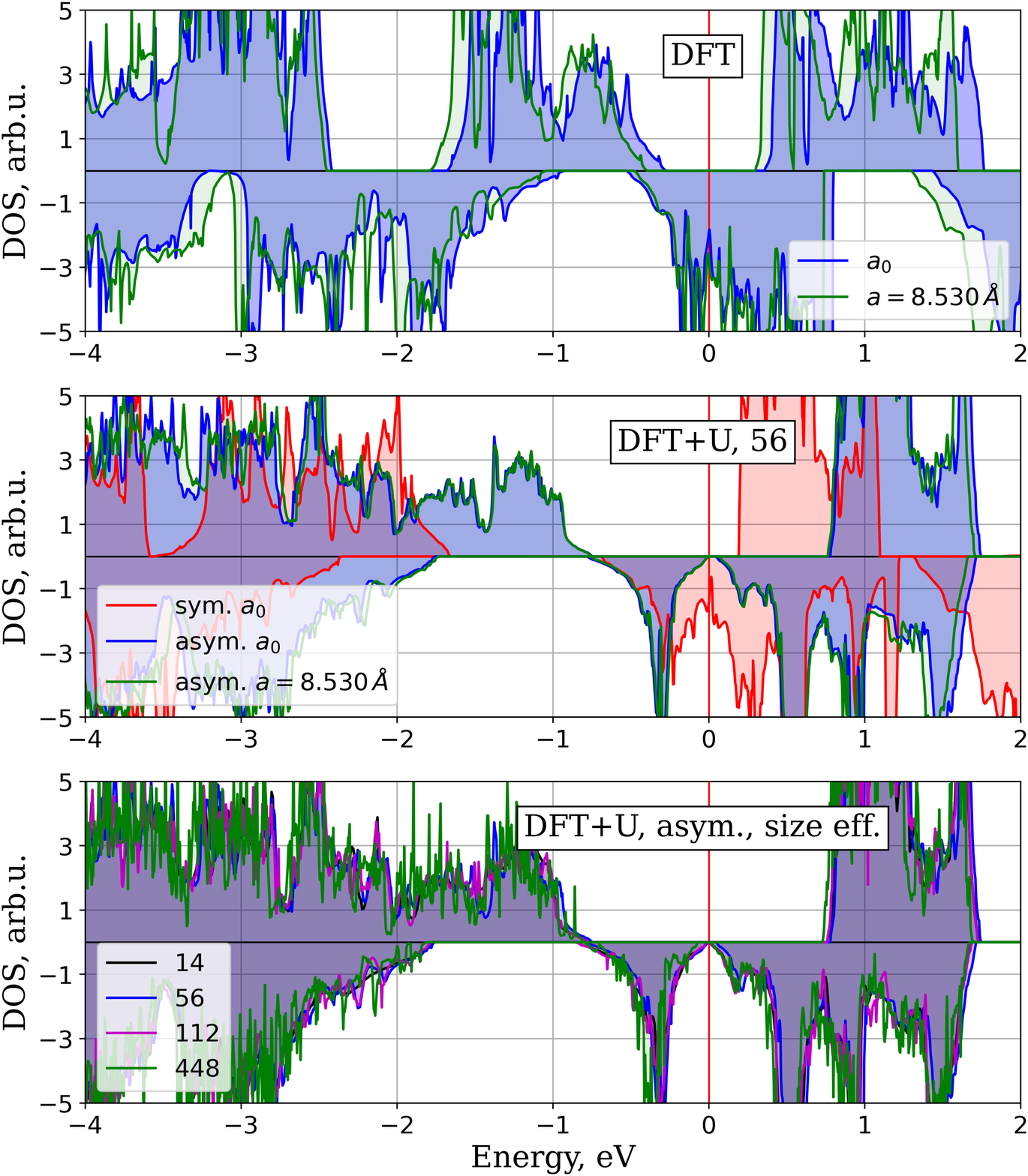}
			\end{center}
		\end{minipage}
		\caption{The total density of states for the cubic phase of magnetite.}
		\label{fig_dos_def_free}
	\end{figure}

	The total density of states (DOS) for the spin up and the spin down electrons obtained for the defect-free magnetite cubic phase in different approximations are shown in Figure~\ref{fig_dos_def_free}. As the same initial approximations to the wave function and the electron density is used, there is no sufficient differences between the DOS calculated at the equilibrium and non-equilibrium (${a=8.530\,\angstrom}$) lattice constants. DFT+U with the symmetry constraint on the electron density and on the wave function does not predict a band gap in the cubic phase of magnetite. The supercell size effect is shown in the bottom of Figure~\ref{fig_dos_def_free}. In this work a single pattern of charge-orbital ordering is considered, which has the lowest total energy (the case 'm2 -545' in \cite{shutikova_vacancy_2021}). 
	
\begin{figure}[]
		\begin{minipage}[]{0.99\linewidth}
			\begin{center}
				\includegraphics[width=0.55\linewidth]{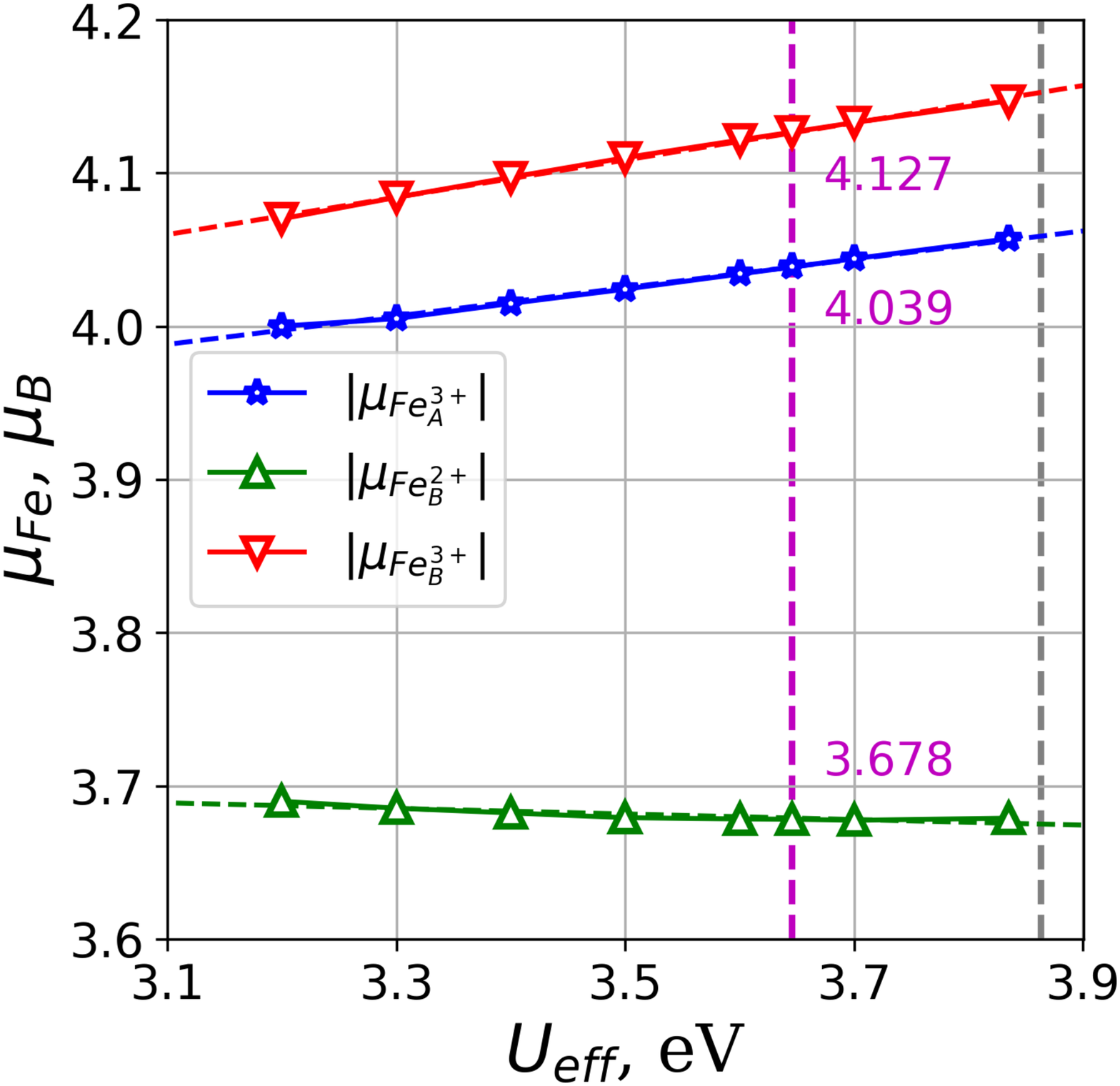}
			\end{center}
		\end{minipage}
		\caption{The $U_{\mathrm{eff}}$ effect on the magnitude of the cation magnetic moments for the assymetric ground state.}
		\label{fig_mag_Ueff}
	\end{figure}

	The total magnetic moment $\mu_{\mathrm{tot}}=4.0\,\mu_{\mathrm{B}}$/f.u. is found in all ferrimagnetic defect-free supercells. The atomic relaxation at a fixed lattice constant do not change $\mu_{\mathrm{tot}}$. In pure DFT without taking into account strong electronic correlations ($U_{\mathrm{eff}}=0$), there are no differences between Fe$_{\mathrm{B}}^{2+}$ and Fe$_{\mathrm{B}}^{3+}$. In DFT+U without the symmetry constraint on the electron density and on the wave function the difference between Fe$_{\mathrm{B}}^{2+}$ and Fe$_{\mathrm{B}}^{3+}$ exists and increases with $U_{\mathrm{eff}}$ increasing (see Figure~\ref{fig_mag_Ueff}). 
	
	There are small stress tensor components in the defect-free cubic supercell with ${N_{\mathrm{at}}=56}$: ${\sigma_{\mathrm{xx}}=-3.25}$, ${\sigma_{\mathrm{yy}}=-3.38}$, ${\sigma_{\mathrm{zz}}=6.56}$, ${\tau_{\mathrm{xy}}=1.50}$, ${\tau_{\mathrm{yz}}=-0.06}$, and ${\tau_{\mathrm{xz}}=-0.07}$ kbar, while the total external pressure of the supercell is close to zero (${U_{\mathrm{eff}}=3.5}$~eV). This result may be attributed to the symmetry constraint on the ionic configuration in calculation of the equilibrium (zero pressure) lattice constant.
	
	\textbf{Geometry optimization of interstitial configurations}
	
\begin{figure}[]
		\begin{minipage}[]{0.99\linewidth}
			\begin{center}
				\includegraphics[width=0.45\linewidth]{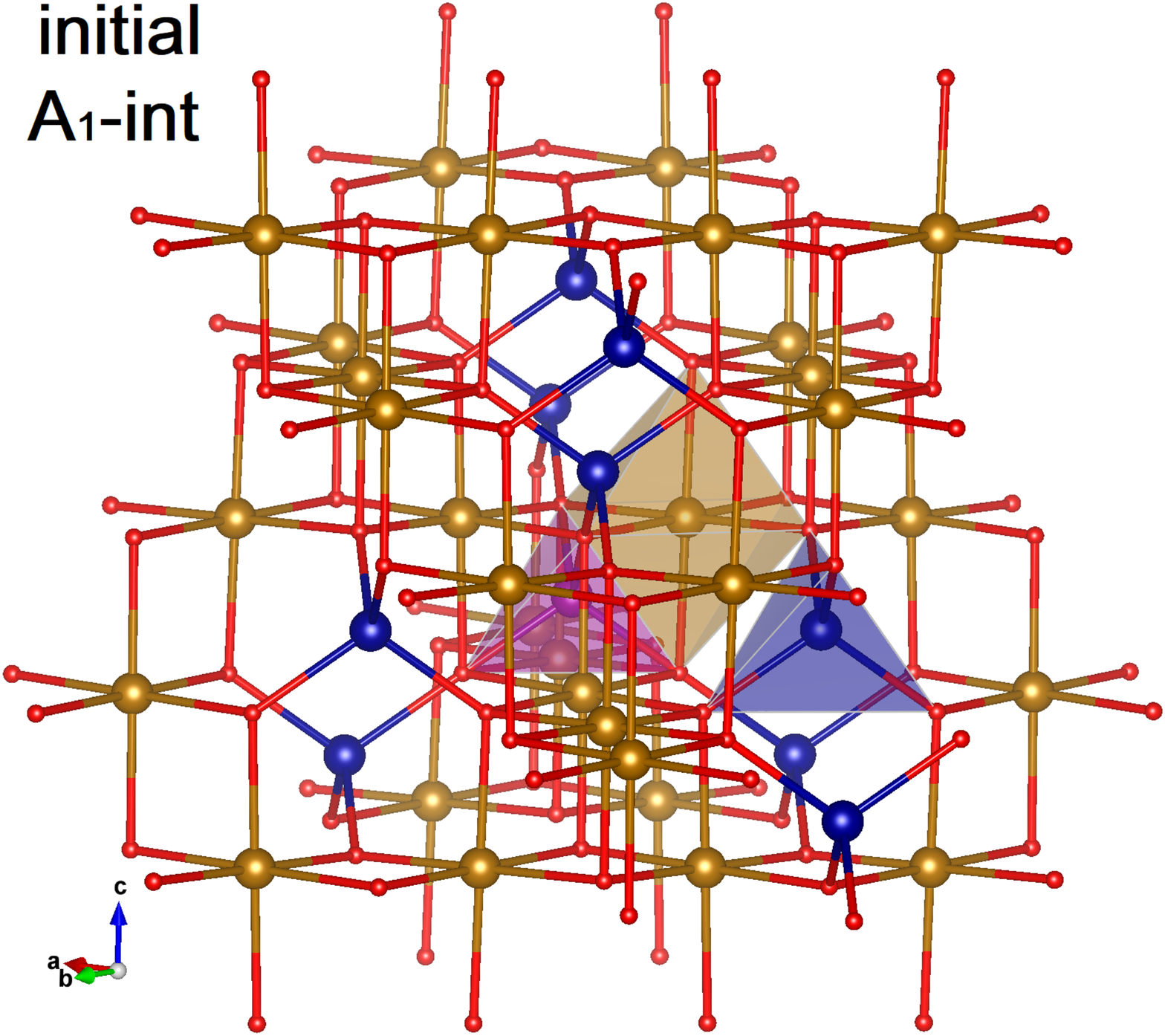}
				\includegraphics[width=0.45\linewidth]{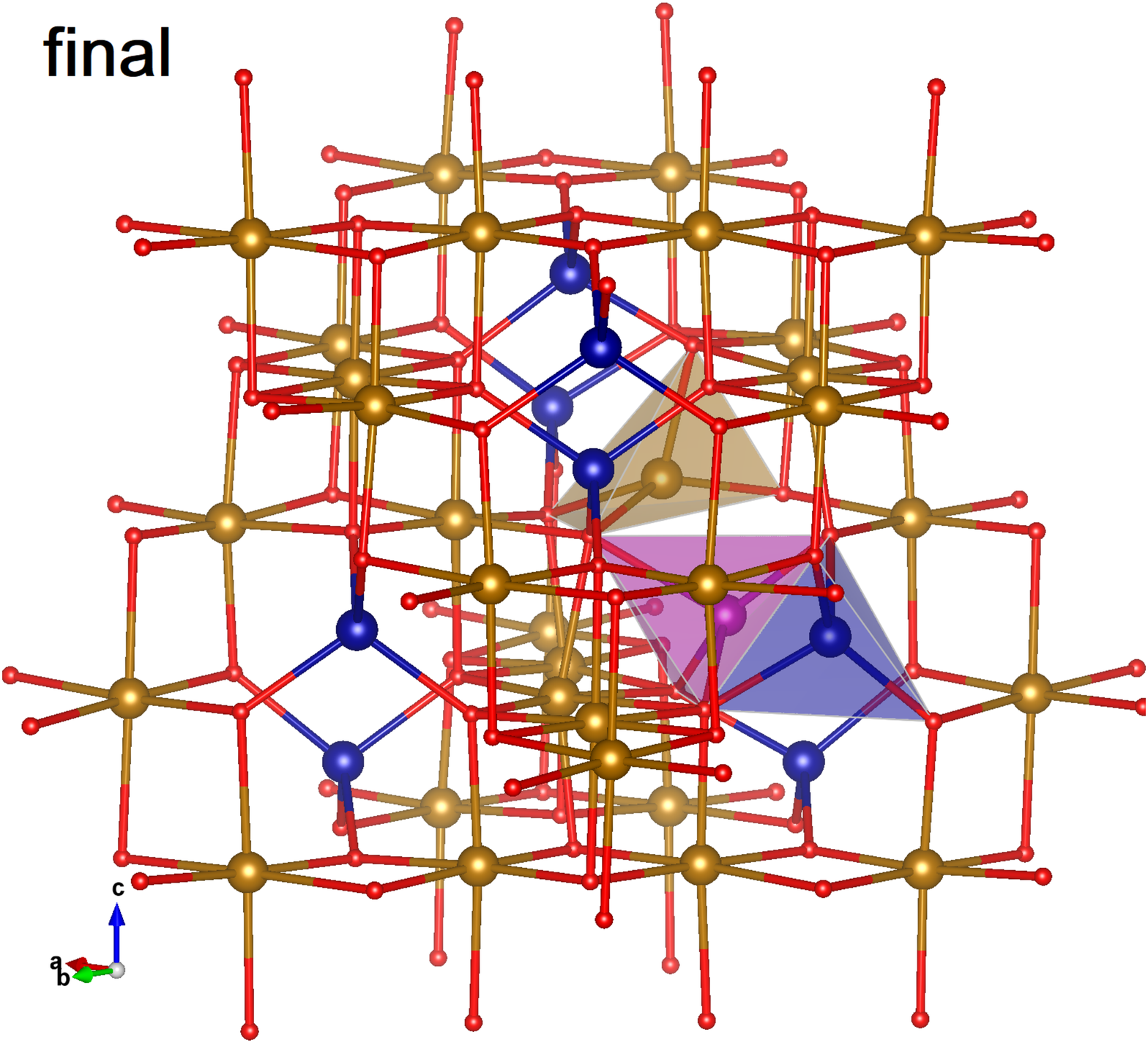}
				\includegraphics[width=0.45\linewidth]{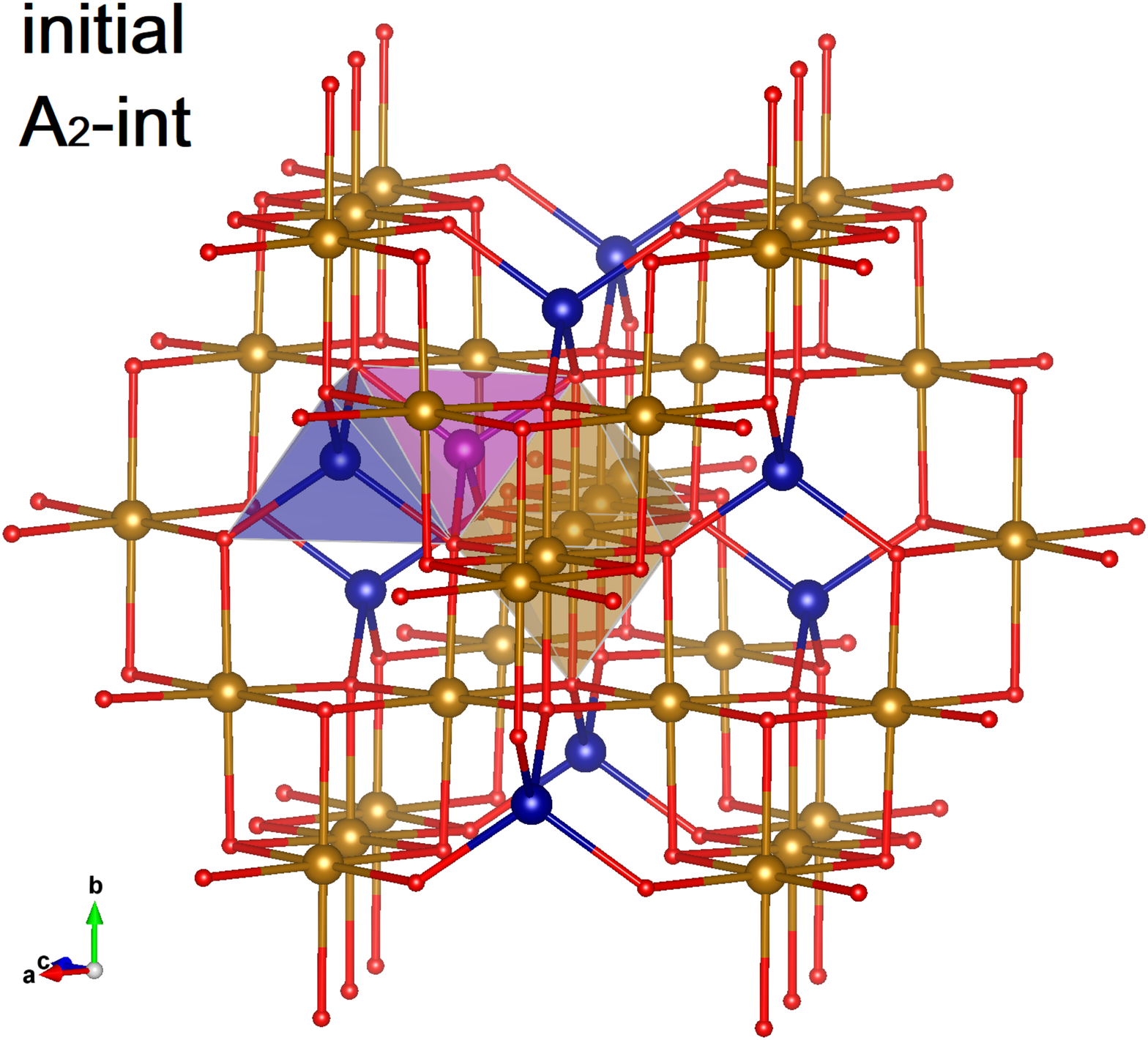}
				\includegraphics[width=0.45\linewidth]{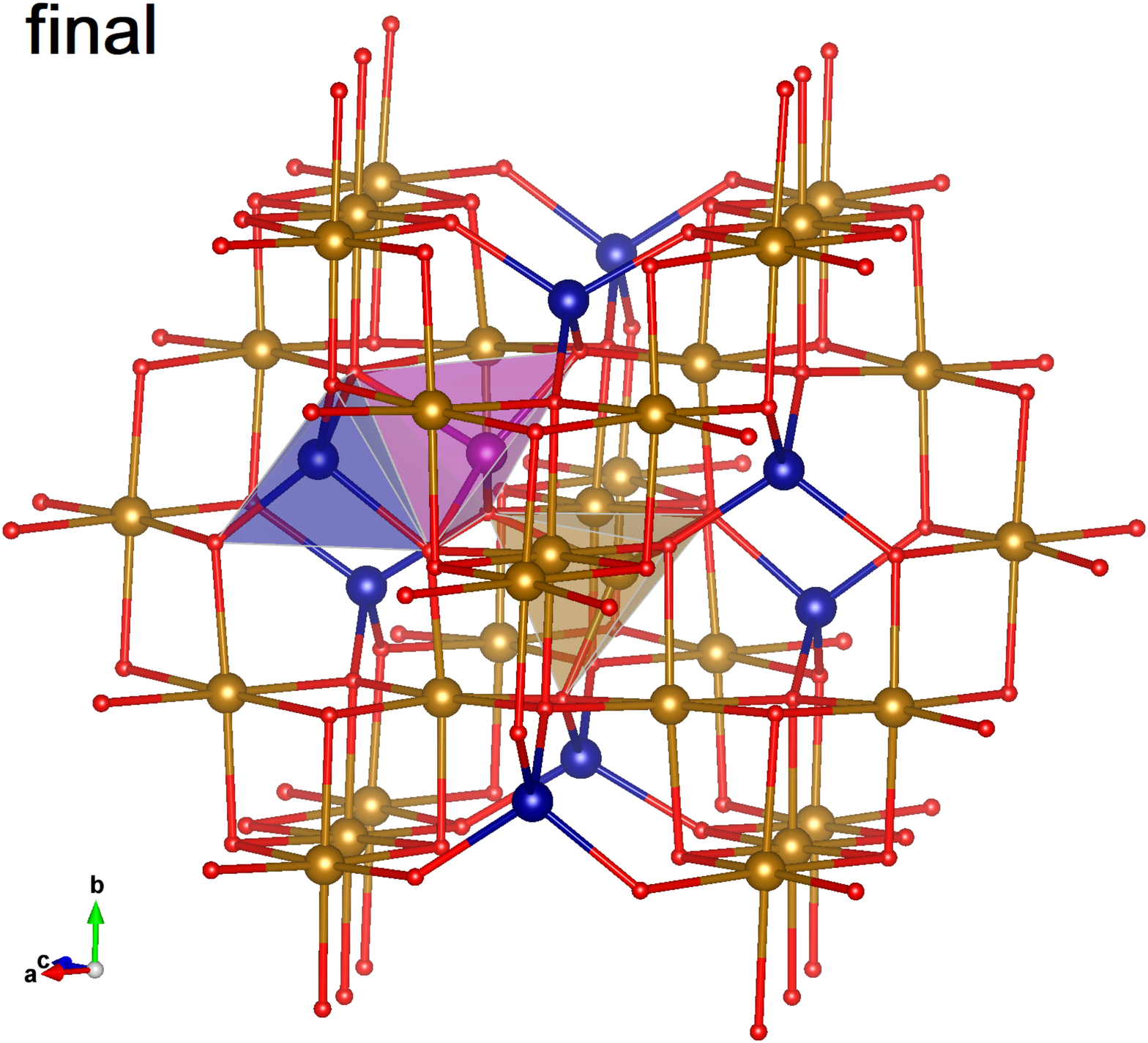}
				\includegraphics[width=0.45\linewidth]{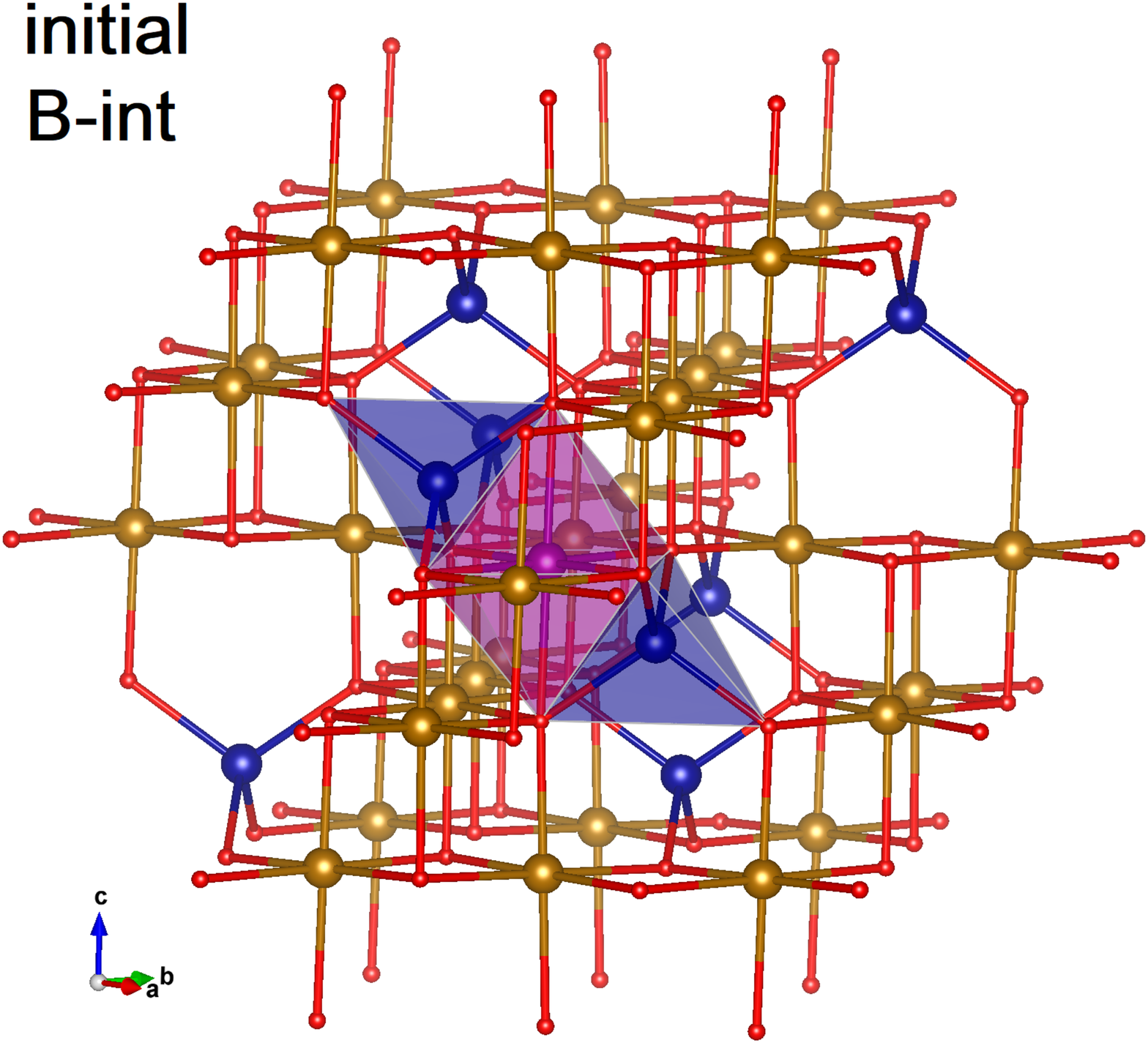}
				\includegraphics[width=0.45\linewidth]{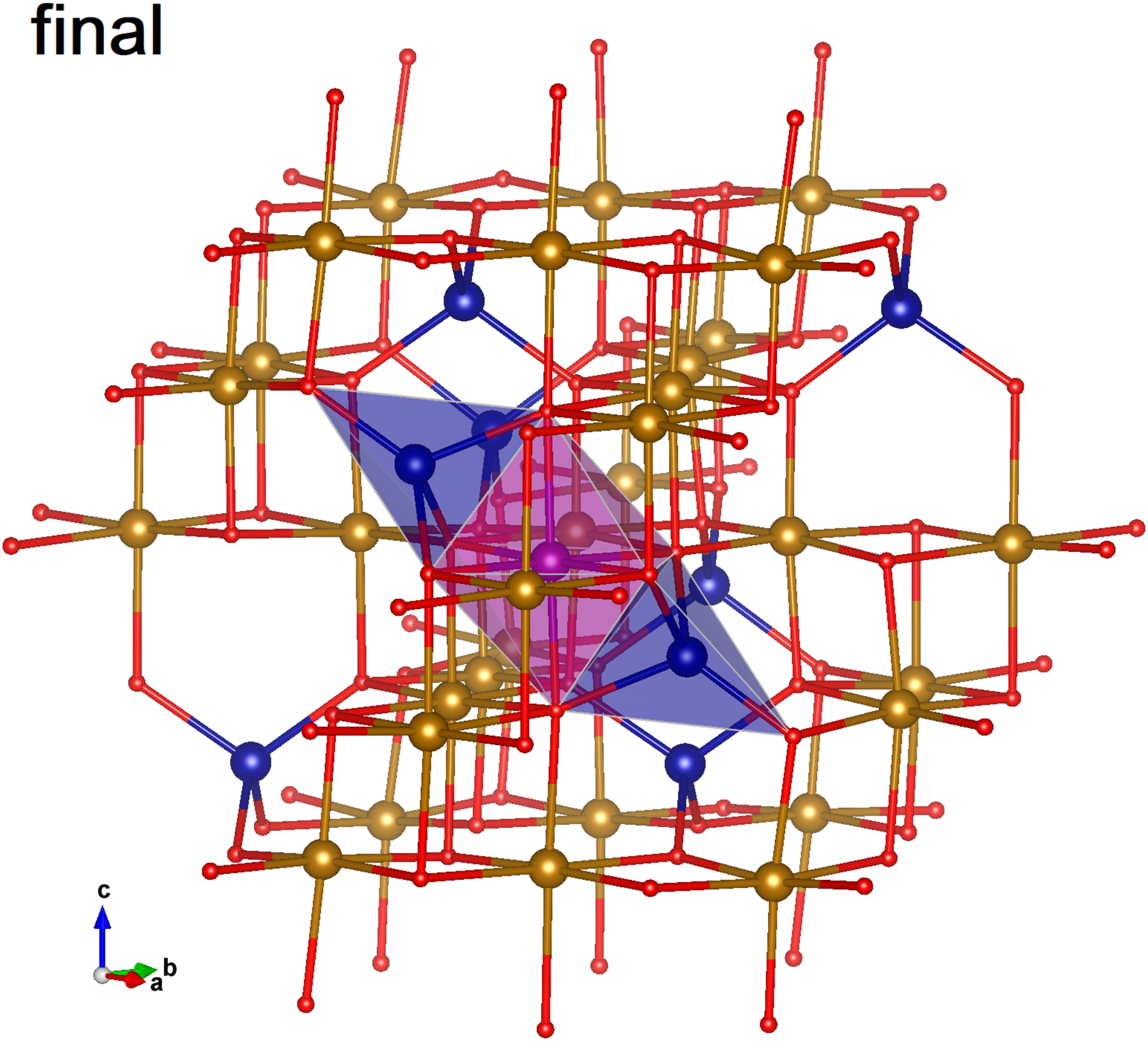}
			\end{center}
		\end{minipage}
		\caption{Initial and final geometries of the defects with the lowest formation energies (see Table~\ref{tabl_int_56}). Fe$_{\mathrm{A}}$, Fe$_{\mathrm{B}}$, and O are shown as dark-blue, gold, and red spheres. The defect cation and its polyhedron are shown in pink.}
		\label{fig_int56_geom}
	\end{figure}

	The properties of the local minima set obtained after geometry optimization at a fixed lattice constant in DFT and DFT+U with ${U_{\mathrm{eff}}=3.5}$~eV are given in Table~\ref{tabl_int_56}. Three types of the iron interstitials with the initial positions denoted as A$_1$-int, A$_2$-int, and B-int respectively are considered (see Figure~\ref{fig_int56_geom})~\cite{arras_electronic_2013}. Five initial approximations to the magnetic moment of the defect cation have been applied: ${0.0,\pm4.0,\pm5.0\,\mu_{\mathrm{B}}}$.
	
	\begin{table}[]
		\begin{center}
			\caption{{\small The properties of the local minima set obtained after the geometry optimization at a fixed lattice constant in DFT and DFT+U (${U_{\mathrm{eff}}=3.5}$~eV): the inital approximation to the magnetic moment of an interstitial cation ($\mu_{{\mathrm{Fe}}_{\mathrm{int}}}^0$, $\mu_{\mathrm{B}}$), the total energy difference (${\Delta E=E_{\mathrm{def}}-E_0}$, eV), the defect formation energy ($E^{\mathrm{f}}_{\mathrm{def}}$, eV), the total magnetic moment difference ($\Delta \mu_{\mathrm{tot}}=\mu_{\mathrm{def}}-\mu_0$, $\mu_{\mathrm{B}}$), the final magnetic moment of the interstitial cation ($\mu_{{\mathrm{Fe}}_{\mathrm{int}}}$, $\mu_{\mathrm{B}}$), the total external pressure difference ($\Delta p=p_{\mathrm{def}}-p_0$, kbar), the displacement of the interstitial atom (from the inital position) ($\Delta r$, $\angstrom$)}}
			\begin{tabular}{ccccccccc}
				\hline  
				initial site & $\mu_{{\mathrm{Fe}}_{\mathrm{int}}}^0$ & $\Delta E$ & $E^{\mathrm{f}}_{\mathrm{def}}$ & $\Delta \mu_{\mathrm{tot}}$ &  $\mu_{{\mathrm{Fe}}_{\mathrm{int}}}$ & $\Delta p$ & $\Delta r$ & \\
				
				\hline
				& $-5.0$ & $-2.724$ & $2.85$* & $+2.0$ & $+3.40$ $\uparrow\uparrow {\mathrm{Fe}}_{\mathrm{B}}$ & $+75.4$ & $0.71$ & \\
				
				& $-$\textbf{4.0} & $-$\textbf{4.674} & \textbf{0.90} & \textbf{+2.0} & \textbf{+3.42} $\uparrow\uparrow {\mathrm{Fe}}_{\mathrm{B}}$ & \textbf{+52.3} & \textbf{2.29} &\\
				
				A$_1$-int & $0.0$ & $-4.628$ &  $0.94$ & $+2.0$ & $+3.79$ $\uparrow\uparrow {\mathrm{Fe}}_{\mathrm{B}}$ & $+61.2$ & $2.15$ &\\
				
				& $+4.0$ & $-2.765$ & $2.81$ & $+2.0$ & $+3.40$ $\uparrow\uparrow {\mathrm{Fe}}_{\mathrm{B}}$ & $+75.0$ & $0.72$ & \\
				
				& $+5.0$ & $-3.473$ & $2.10$ & $-4.0$ & $+3.47$ $\uparrow\uparrow {\mathrm{Fe}}_{\mathrm{B}}$ & $+48.3$& $2.23$ & \\
				\cline{1-8}
				& $-5.0$ & $-3.423$ & $2.15$ & $-10.0$ & $-3.87$ $\uparrow\uparrow {\mathrm{Fe}}_{\mathrm{A}}$ & $+49.4$& $0.30$ &\\
				
				& $-4.0$ & $-4.303$ & $1.27$ & $-8.0$ & $-3.66$ $\uparrow\uparrow {\mathrm{Fe}}_{\mathrm{A}}$ & $+37.2$ & $1.60$ & this work\\
				
				A$_2$-int & \textbf{0.0} & $-$\textbf{4.802} &  \textbf{0.77} & \textbf{+2.0} & \textbf{+3.41} $\uparrow\uparrow {\mathrm{Fe}}_{\mathrm{B}}$ & \textbf{+52.9} & \textbf{0.46} &DFT+U  \\
				
				& $+4.0$ & $-1.875$ & $3.70$ & $-6.0$ & $+3.38$ $\uparrow\uparrow {\mathrm{Fe}}_{\mathrm{B}}$ & $+37.2$ & $0.18$ & $U_{\mathrm{eff}}=3.5$~eV\\
				
				& $+5.0$ & $-2.045$ & $3.52$ & $-8.0$ & $+3.42$ $\uparrow\uparrow {\mathrm{Fe}}_{\mathrm{B}}$ & $+41.1$ & $0.43$ &\\
				\cline{1-8}
				& $-5.0$ & $-4.550$ & $1.02$ & $+2.0$ & $+3.61$ $\uparrow\uparrow {\mathrm{Fe}}_{\mathrm{B}}$ & $+39.1$ & $0.02$ &\\
				
				& $-4.0$ & $-5.169$ & $0.40$ & $-6.0$ & $-3.70$ $\uparrow\uparrow {\mathrm{Fe}}_{\mathrm{A}}$ & $+39.5$ & $0.01$ &\\
				
				B-int & \textbf{0.0} & $-$\textbf{5.246} &  \textbf{0.32} & $-$\textbf{6.0} & $-$\textbf{3.73} $\uparrow\uparrow {\mathrm{Fe}}_{\mathrm{A}}$ & \textbf{+41.4} &\textbf{0.01}& \\
				
				& $+4.0$ & $-5.238$ & $0.33$ & $-6.0$ & $-3.71$ $\uparrow\uparrow {\mathrm{Fe}}_{\mathrm{A}}$& $+40.5$ & $0.00$ & \\
				
				& $+5.0$ & $-4.016$ & $1.55$ & $-8.0$ & $-3.70$ $\uparrow\uparrow {\mathrm{Fe}}_{\mathrm{A}}$ & $+34.4$ & $0.01$& \\
				\hline	
				B-int	& \begin{tabular}{c}
					$-5.0$ \\
					$-4.0$ \\
					$0.0$ \\
				\end{tabular} & $-5.098$ & $3.14$** & $-6.0$ & $-3.41$ $\uparrow\uparrow {\mathrm{Fe}}_{\mathrm{A}}$ & $+32.8$ & $0.00$ & this work \\
				
				& \begin{tabular}{c}
					$+4.0$ \\
					$+5.0$ \\
				\end{tabular} & $-4.302$ & $3.94$ & $-2.0$ & $+3.28$ $\uparrow\uparrow {\mathrm{Fe}}_{\mathrm{B}}$ & $+16.11$ & $0.00$ & DFT\\
				
				\hline	
				
				A-int & $-$ & $-$ & $5.40$ & $+2.0$ & $\uparrow\uparrow {\mathrm{Fe}}_{\mathrm{B}}$ & $-$ & $-$ & DFT \cite{hendy_ab_2003}\\
				
				A-int & $-$ & $-$ & $1.25$ & $-$ & $-$ & $-$ & $-$ & DFT+U  \cite{li_influence_2016}\\
				
				A$_1$-int & $-$ & $-$ & $-$ & $+2.0$ & $+3.13$ $\uparrow\uparrow {\mathrm{Fe}}_{\mathrm{B}}$ & $-$ & $-$ & DFT+U \cite{arras_electronic_2013}\\
				
				A$_2$-int & $-$ & $-$ & $-$ & $-4.0$ & $-3.09$ $\uparrow\uparrow {\mathrm{Fe}}_{\mathrm{A}}$ & $-$ & $-$ & DFT+U \cite{arras_electronic_2013}\\
				
				\hline
				
				& $-$ & $-$ & $3.26$ & $-6.0$ & $\uparrow\uparrow {\mathrm{Fe}}_{\mathrm{A}}$ & $-$ & $-$ & DFT \cite{hendy_ab_2003}\\
				
				B-int & $-$ & $-$ & $0.81$ & $-$ & $-$ & $-$ & $-$ & DFT+U \cite{li_influence_2016}\\
				
				& $-$ & $-$ & $-$ & $-6.0$ & $-3.45$ $\uparrow\uparrow {\mathrm{Fe}}_{\mathrm{A}}$ & $-$ & $-$ & DFT+U \cite{arras_electronic_2013} \\
				
				\hline
				\multicolumn{9}{c}{* $E_{\mathrm{at}}=-5.57$~eV in DFT+U with $U_{\mathrm{eff}}=3.5$~eV} \\
				\multicolumn{9}{c}{** $E_{\mathrm{at}}=-8.24$~eV in DFT without the Hubbard U} \\
			\end{tabular}
			\label{tabl_int_56}
		\end{center}	
	\end{table}
	
	The octahedral interstitial (B-int case) has the lowest formation energy in DFT+U (Table~\ref{tabl_int_56}). In this case the interstitial atom does not displace from its initial position after the geometry optimization. In A$_1$-int and A$_2$-int cases, on the contrary, the interstitial atom displacements from the initial position (see $\Delta r$ Table~\ref{tabl_int_56}) are found to be comparable to the distance $d$(Fe$_{\mathrm{B}}$-O)=2.08~$\angstrom$ in the defect free supecell. The distances $d$(Fe$^{\mathrm{def}}$-Fe$^1_{\mathrm{A}}$), $d$(Fe$^{\mathrm{def}}$-Fe$^1_{\mathrm{B}}$) of 2.46, 2.42~$\angstrom$, and 2.44, 2.51~$\angstrom$ in the A$_1$-int, and A$_2$-int initial cases are smaller than $d$(Fe$_{\mathrm{B}}$-Fe$_{\mathrm{B}}$)=3.00~$\angstrom$ in the lattice without defects. Thus, the defects obtained after the geometry optimization of the A$_1$-int and A$_2$-int cases seem not to be the isolated defects.
	
	It is very difficult to make a prediction about the best initial approximations for a wave function and a charge density in calculations of the defects. For example, the initial magnetic moments of $+4.0$ and $-5.0$~$\mu_{\mathrm{B}}$ in A$_1$-int case give very close minima, as well as $0.0$ and $\pm4.0$~$\mu_{\mathrm{B}}$ in B-int case. In this work we have not found the local minimum with the total magnetic moment difference of $-4.0$~$\mu_{\mathrm{B}}$ reported in~\cite{arras_electronic_2013}.
	
\begin{figure}[]
		\begin{minipage}[]{0.99\linewidth}
			\begin{center}
				\includegraphics[width=0.49\linewidth]{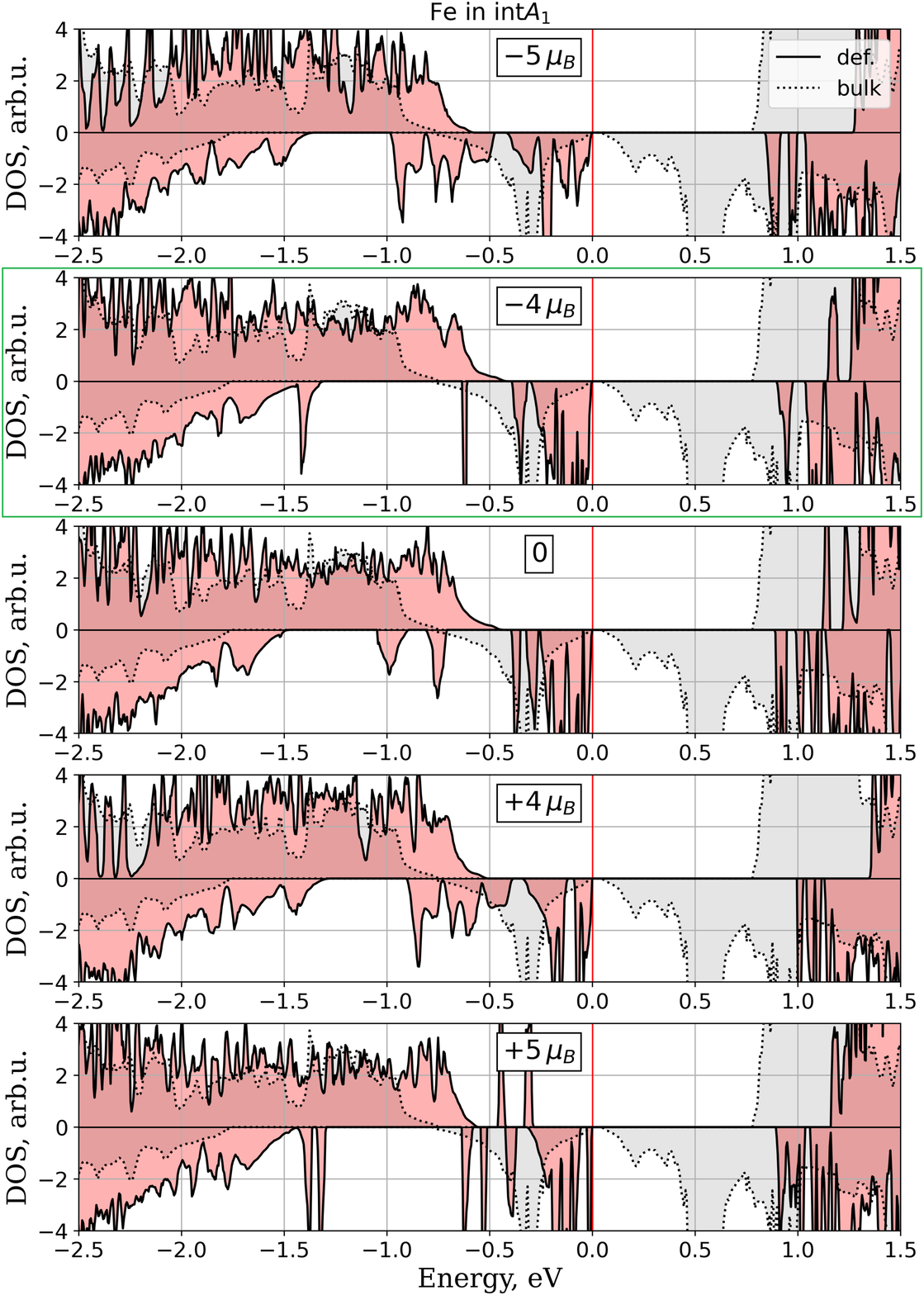}
				\includegraphics[width=0.49\linewidth]{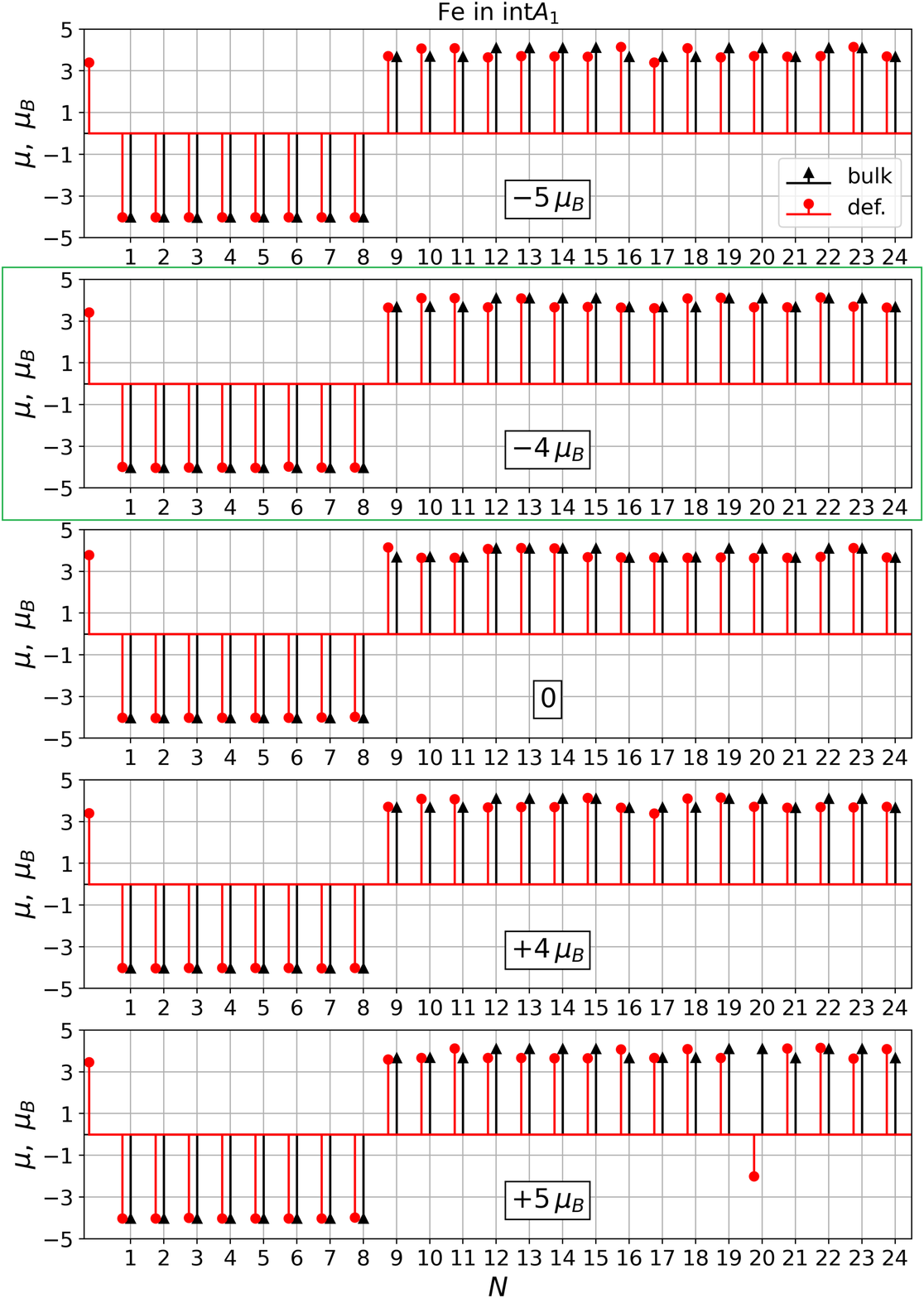}
			\end{center}
		\end{minipage}
		\caption{The total density of state and the magnetic moments of cations for the A$_1$-int initial position (see Table~\ref{tabl_int_56}). The defect cation, A-, and B-cation numbers are denoted as $N=0$, $N=1\div8$, and $N=9\div24$, respectively. The magnetic moments of defect free bulk are shown in black for comparison. The lowest energy local minimum is marked by a green frame.}
		\label{fig_int56_dos_mag_A1}
	\end{figure}

\begin{figure}[]
		\begin{minipage}[]{0.99\linewidth}
			\begin{center}
				\includegraphics[width=0.49\linewidth]{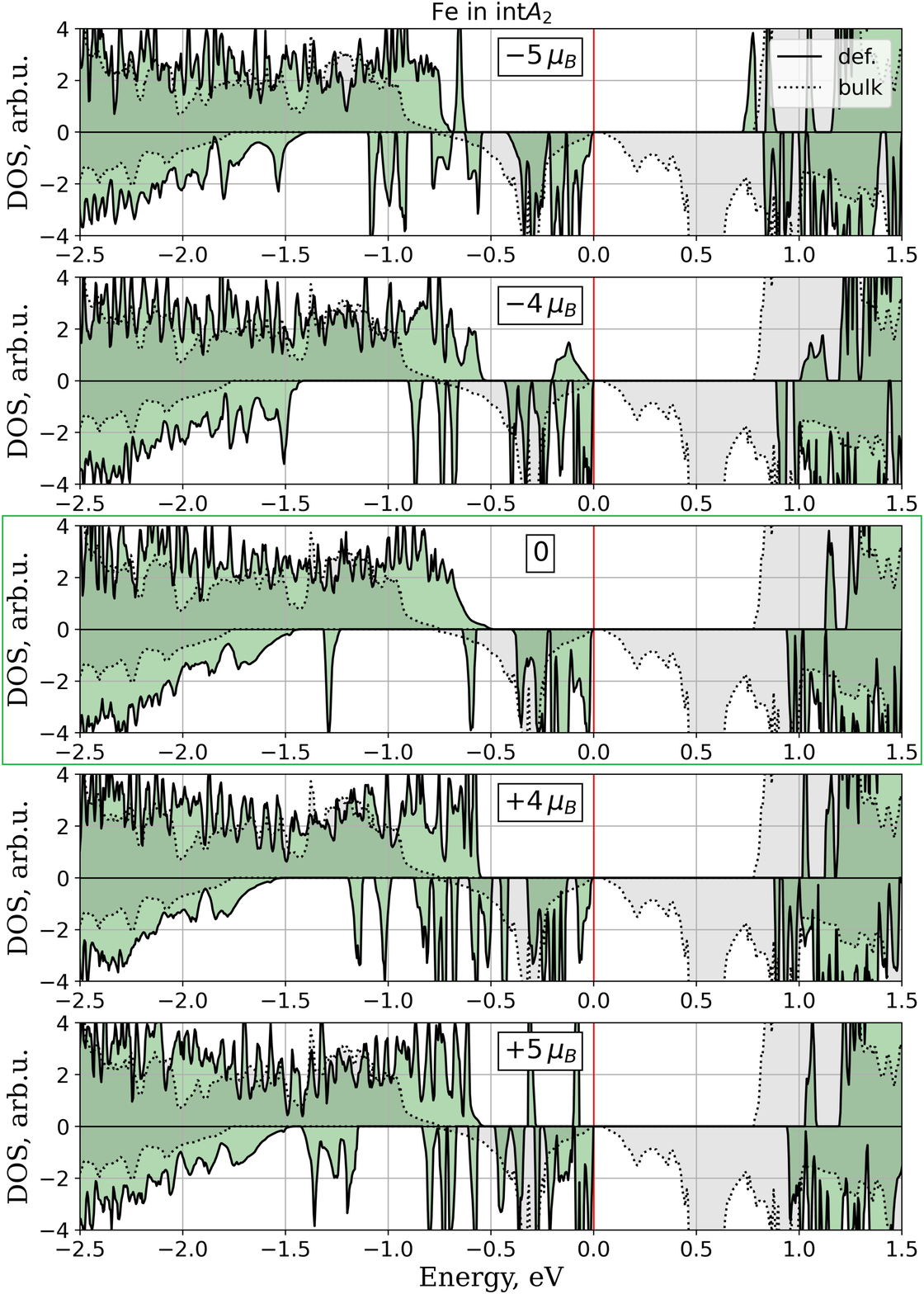}
				\includegraphics[width=0.49\linewidth]{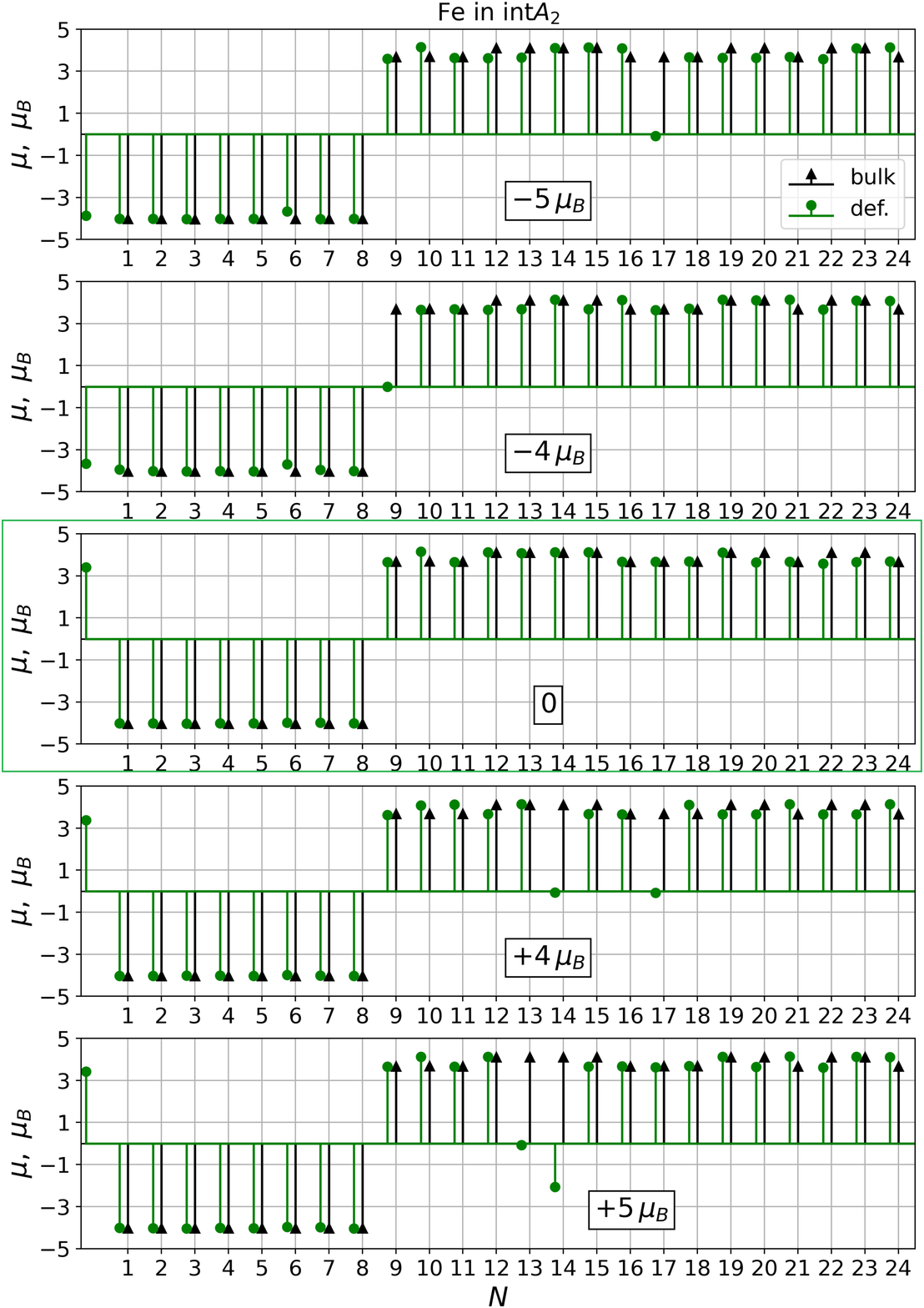}
			\end{center}
		\end{minipage}
		\caption{The total density of states and the magnetic moments of cations for the A$_2$-int initial position (see Table~\ref{tabl_int_56}).}
		\label{fig_int56_dos_mag_A2}
	\end{figure}

\begin{figure}[]
		\begin{minipage}[]{0.99\linewidth}
			\begin{center}
				\includegraphics[width=0.49\linewidth]{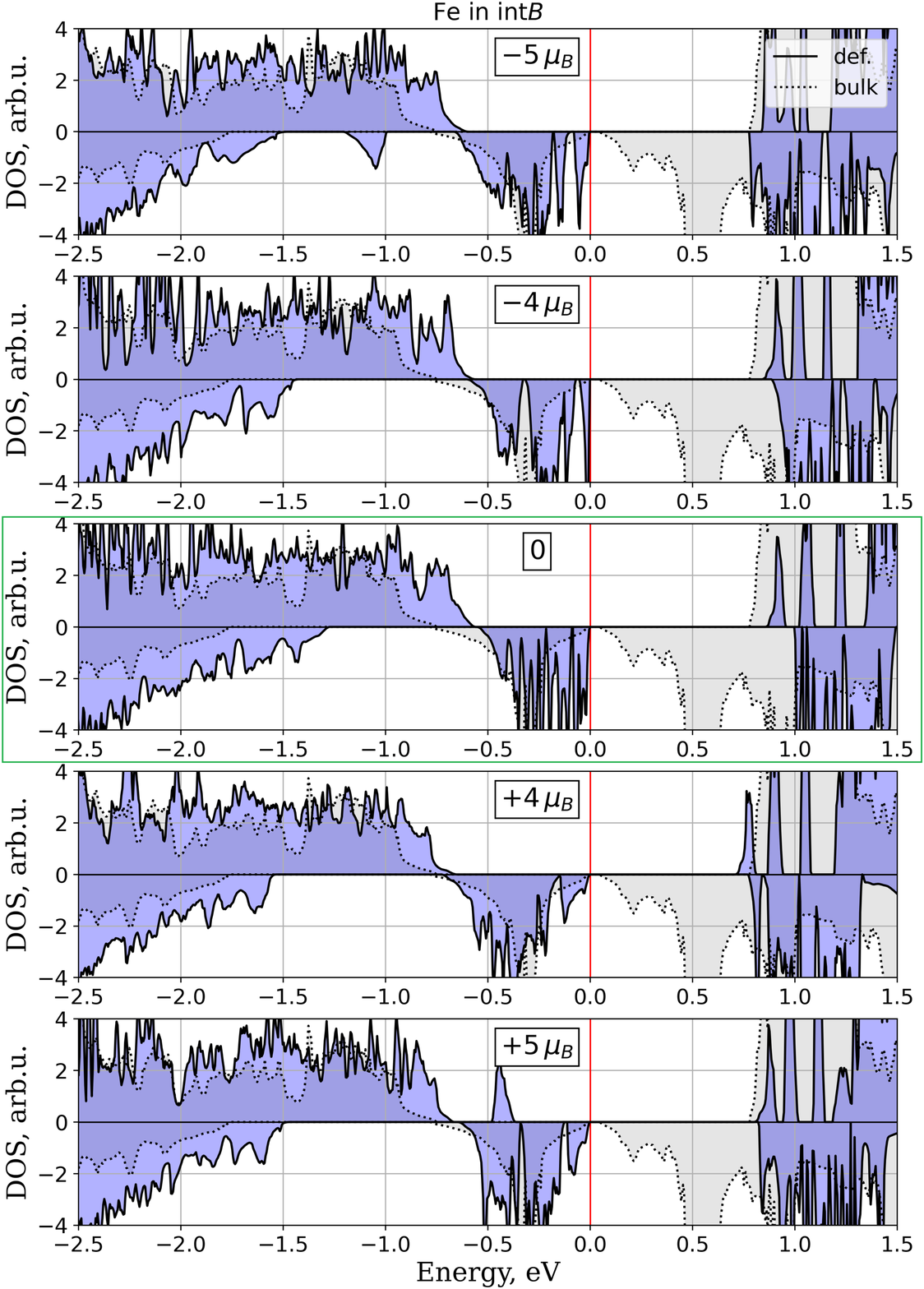}
				\includegraphics[width=0.49\linewidth]{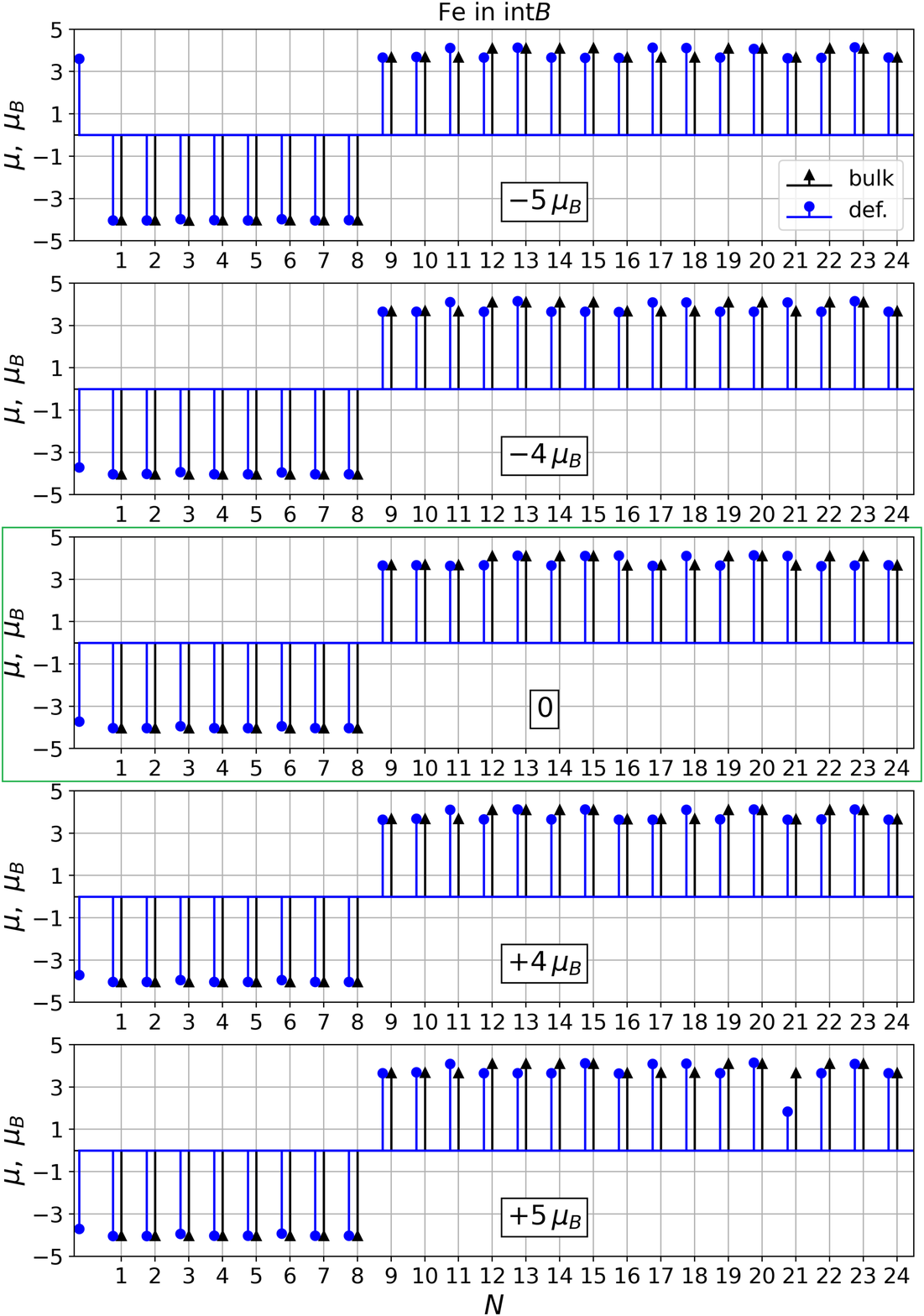}
			\end{center}
		\end{minipage}
		\caption{The total density of states and the magnetic moments of cations for the B-int initial position (see Table~\ref{tabl_int_56}).}
		\label{fig_int56_dos_mag_B}
	\end{figure}

	The DOS and the cation magnetic moments of the DFT+U cases in Table~\ref{tabl_int_56} are shown in Figure~\ref{fig_int56_dos_mag_A1}, Figure~\ref{fig_int56_dos_mag_A2}, and Figure~\ref{fig_int56_dos_mag_B}. In Table~\ref{tabl_int_56} one can see the relatively large, compared to the minimum value, formation energies obtained in the $+4.0,+5.0\,\mu_{\mathrm{B}}$ cases in the A$_2$-int initial position. At the same time, sufficient changes in the magnetic moment of two B-cations are observed in these cases (Figure~\ref{fig_int56_dos_mag_A2}). The similar considerations may be given in the other cases: there are the relatively large, compared to the minimum value, defect formation energies and a sufficient change in the magnetic moment of one B-cation in the $+5.0\,\mu_{\mathrm{B}}$ case for the A$_1$-int initial position, in the $-4.0,\,-5.0\,\mu_{\mathrm{B}}$ cases for the A$_2$-int initial position, and in the $+5.0\,\mu_{\mathrm{B}}$ case for the B-int initial position.
	
\begin{figure}[]
		\begin{minipage}[]{0.99\linewidth}
			\begin{center}
				\includegraphics[width=0.49\linewidth]{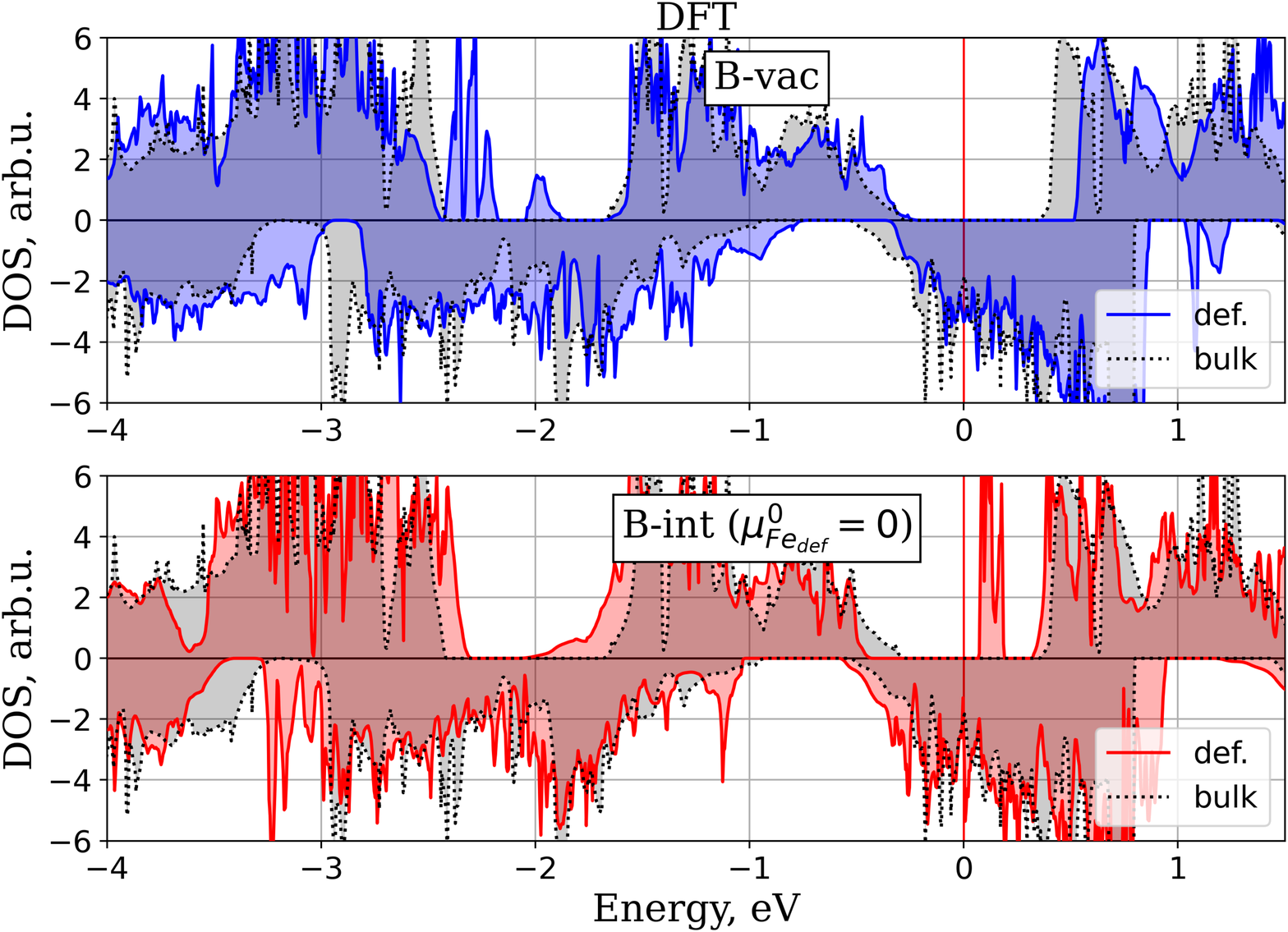}
				\includegraphics[width=0.49\linewidth]{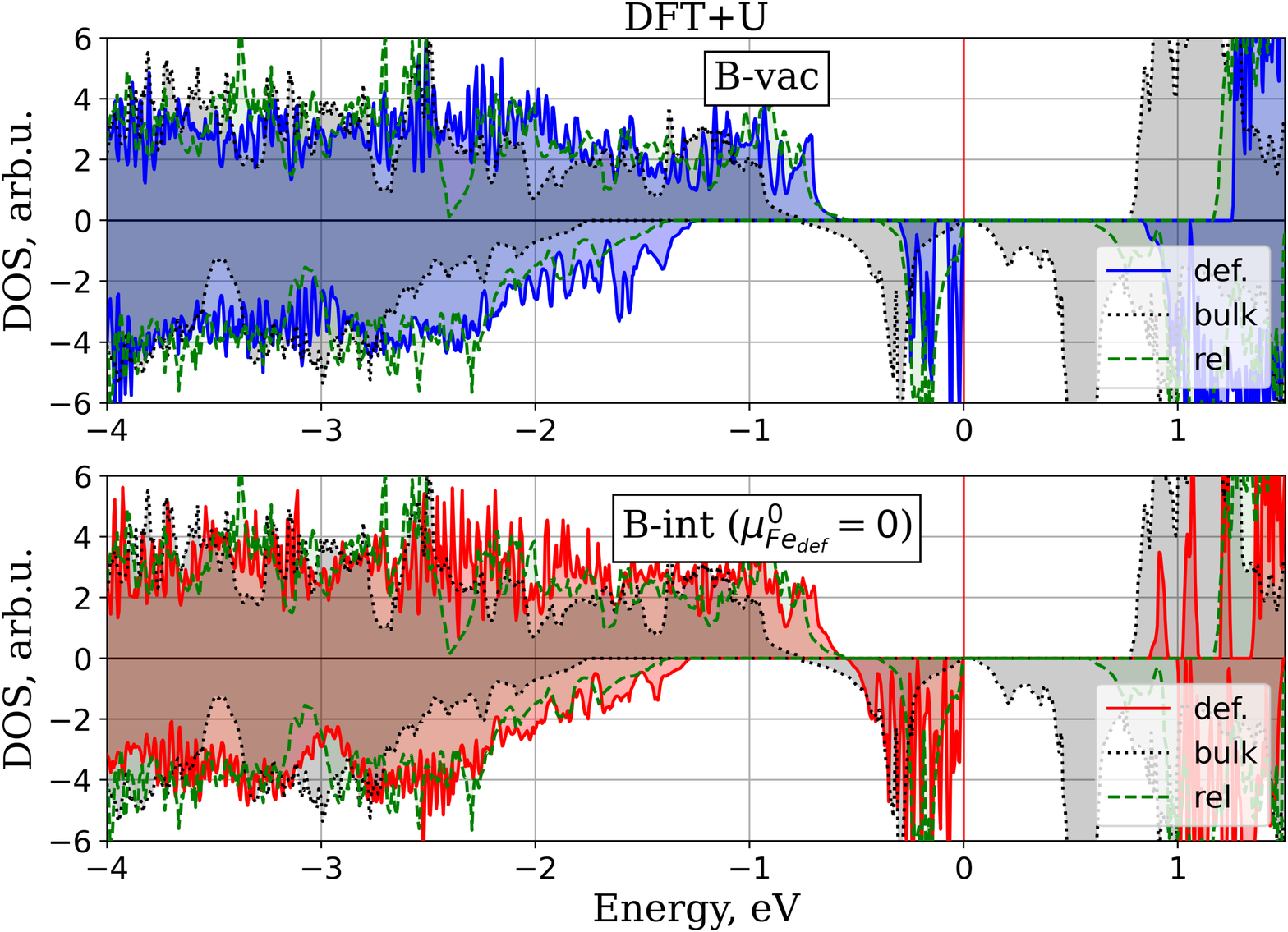}
			\end{center}
		\end{minipage}
		\caption{The total density of states of the B-vacancy and the B-interstitial in DFT and DFT+U.}
		\label{fig_dos_56_def}
	\end{figure}

	The DOS of the B-vacancy and the B-interstitial calculated in DFT and DFT+U is shown in Figure \ref{fig_dos_56_def}. The defect formation increases the band gap. This increase is larger than that caused by the atomic relaxation in the defect free supercell.
	
	\textbf{The chemical potential of iron}
	
\begin{figure}[h]
		\begin{minipage}[]{0.99\linewidth}
			\begin{center}
				\includegraphics[width=0.55\linewidth]{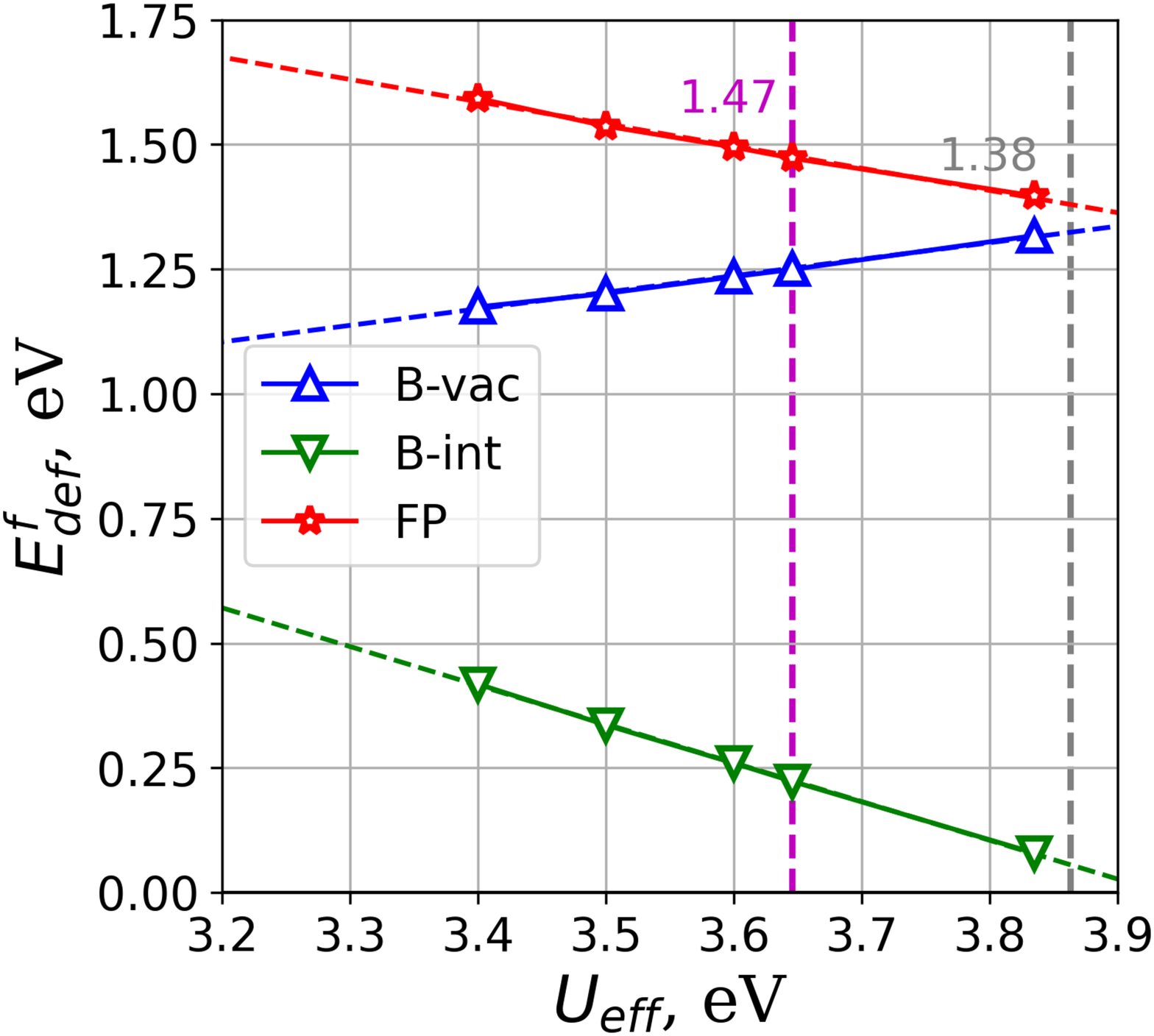}
			\end{center}
		\end{minipage}
		\caption{{\small The effect of $U_{\mathrm{eff}}$ on the defect formation energies.}}
		\label{fig_fp_Ueff2}
	\end{figure}

	The chemical potential of iron $E_{\mathrm{at}}$ is used to estimate the formation energy of the isolated point defect (see the formula~(1) in the main text). All energies in (1) must be calculated within the same approximation of the exchange-correlation functional. The energy per atom in the ferromagnetic body centered cubic phase of iron $E_{\mathrm{bccFe}}$ can be taken as a limit value of $E_{\mathrm{at}}$. However one can obtain two differen values of $E_{\mathrm{bccFe}}$ with and without the Hubbard U (${E_{\mathrm{bccFe}}^{\mathrm{PBE}}=-8.238}$~eV, ${E_{\mathrm{bccFe}}^{\mathrm{PBE+U}}=-5.571}$~eV with ${U_{\mathrm{eff}}=3.5}$~eV). The negative defect formation energies are obtained in DFT+U using $E_{\mathrm{bccFe}}^{\mathrm{PBE}}$. In contrast to the formation energy of the isolated point defect, the formation energy of the Frenkel pair does not depends on $E_{\mathrm{at}}$. This allows comparing the numerical and the experimental results on the Frenkel pair formation energy. The effect of $U_{\mathrm{eff}}$ on the defect formation energies is shown in Figure~\ref{fig_fp_Ueff2}.
	
	\textbf{Supercell size effects}
	
\begin{figure}[h]
		\begin{minipage}[]{0.99\linewidth}
			\begin{center}
				\includegraphics[width=0.485\linewidth]{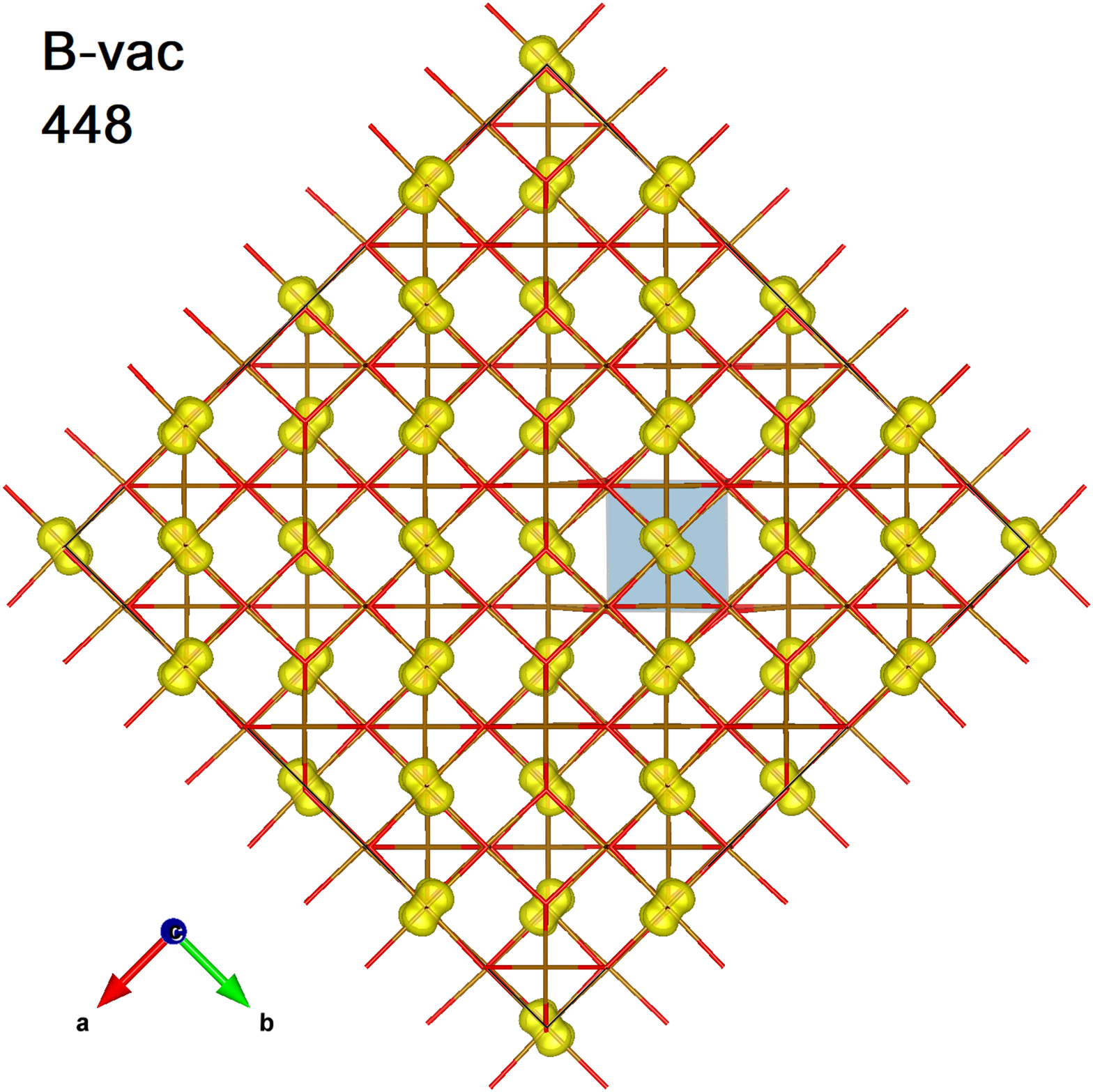}
				\includegraphics[width=0.49\linewidth]{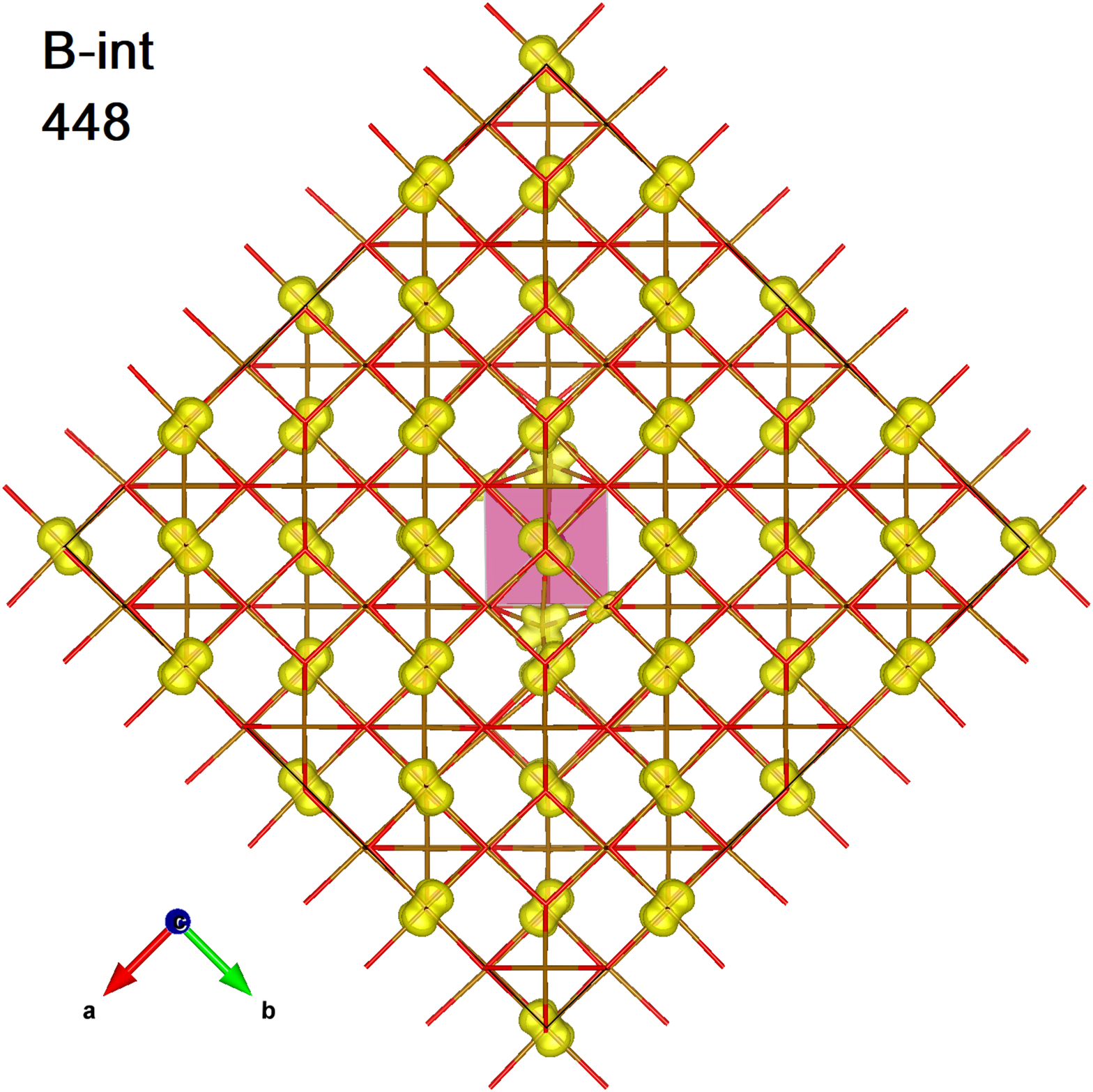}
			\end{center}
		\end{minipage}
		\caption{The optimized geometry and the partial charge density for the B-vac and B-int defects in the DFT+U framework obtained in supecells containing 448 atoms and the defect.}
		\label{fig_pcd_448_def}
	\end{figure}

	The optimized geometry and the partial charge density for the B-vac and B-int defects in the supercell with 448 atoms are shown in Figure~\ref{fig_pcd_448_def}. One can see that the changes induced by the defect formation are local, while the rotation of orbitals induced by the atomic relaxation, cover the whole supercell (see also Figure 15 from the paper \cite{shutikova_vacancy_2021}).
	
